\documentclass[journal]{IEEEtran}
\usepackage[noadjust]{cite}
\usepackage{srcltx}
\usepackage{graphicx}
\usepackage{epstopdf}
\usepackage{psfrag}
\usepackage{epsfig}
\usepackage[tbtags,sumlimits,nointlimits,reqno]{amsmath}
\usepackage{amssymb}
\usepackage{xcolor}
\usepackage{float}
\usepackage{subfigure}
\usepackage[most]{tcolorbox}
\usepackage{soul}
\usepackage{footnote}
\usepackage{amsmath}
\usepackage{color}
\newcommand{\jstyle}[1]{\textit{\textbf{{1}}}}
\newcommand{\rstyle}[1]{\textit{\textbf{{1}}}}
\newcommand{\cstyle}[1]{\textit{\textbf{{1}}}}
\newcommand{\bstyle}[1]{\textit{\textbf{{1}}}}
\newcommand{\istyle}[1]{\emph{\textbf{{1}}}}
\newcommand{\tbpstyle}[1]{\textit{{1}}}
\usepackage[framemethod=TikZ]{mdframed}
\usepackage{lipsum}
\mdfdefinestyle{MyFrame}{%
	linecolor=blue,
	outerlinewidth=2pt,
	roundcorner=20pt,
	innertopmargin=\baselineskip,
	innerbottommargin=\baselineskip,
	innerrightmargin=20pt,
	innerleftmargin=20pt,
	backgroundcolor=gray!50!white}

\begin{document}
	\title{On-Chip Nonreciprocal Josephson Frequency Multiplier for Multiplexed Control of Superconducting Qubits}
	\author{Sajjad Taravati\\
		 School of Electronics and Computer Science, University of Southampton, Southampton SO17 1BJ, UK\\e-mail: s.taravati@soton.ac.uk
	%	\thanks{}%
	%	\thanks{}
	}
	% The paper headers
	\markboth{}%
	%\markboth{}
	{* \MakeLowercase{\textit{et al.}}: Bare Demo of IEEEtran.cls for IEEE Journals}

	\maketitle

\textbf{Abstract:} This paper presents an on-chip nonreciprocal space-time-periodic Josephson frequency multiplier that up-converts a single low-frequency input tone into a comb of high-order harmonics while enforcing directional signal propagation along the bus. We show that the harmonic content and bandwidth of the generated comb can be controlled through the applied DC and RF flux bias, and that individual output channels can be selectively addressed by tuning the bus operating point. We characterize the resulting nonreciprocal isolation between channels and discuss the associated design trade-offs, including the influence of modulation depth on harmonic-generation efficiency and back-reflection. The proposed device delivers multiple mutually isolated frequency channels from a single low-frequency source and it is well suited to applications requiring multiplexed delivery of tones to spatially distributed, frequency-addressable loads; we discuss multiplexed control of superconducting qubits as one such potential application area. Master-equation simulations of a four-qubit array driven by the harmonic comb, together with a one-dimensional line model of the directional conversion, quantify the resulting crosstalk and directional asymmetry; all results are simulations, and no experimental demonstration is claimed. The proposed architecture offers a compact, integrated route toward on-chip frequency multiplexing and directional isolation for cryogenic microwave systems.

	% Note that keywords are not normally used for peerreview papers.

\IEEEpeerreviewmaketitle
	
\section{Introduction}\label{sec:introduction}
Frequency conversion is a valuable function as it enables a single low-frequency input to be up-converted into one or more high-frequency output tones without dedicated high-frequency source electronics for each tone. Superconducting frequency converters are attractive for this purpose due to their ultra-low loss and high coherence at millikelvin temperatures, and they have found application in quantum sensing~\cite{berlin2020axion,dixit2021searching,bass2024quantum}, hybrid quantum networks~\cite{han2021microwave}, and control of quantum systems that require faithful signal transfer across disparate frequency bands~\cite{albrecht2014waveguide,bock2018high,maring2018quantum,santiago2022resonant,taravati2024efficient,geus2024low}. Combining frequency conversion with nonreciprocity is especially useful: a directional, frequency-selective bus can deliver signals to distributed loads while simultaneously isolating each load from reflections, back-propagating noise, and cross-coupling from neighboring channels~\cite{gambetta2017building,he2022control}.

One motivating context for such a device is the cryogenic control of superconducting qubits, where the conventional architecture requires dedicated coaxial lines for flux tuning (Z-control), microwave drive (XY-control), and readout for every few qubits~\cite{clarke2008superconducting,kjaergaard2020superconducting,anferov2024superconducting}. This results in substantial cryogenic wiring burden as qubit count increases, with each line contributing thermal load and signal attenuation, and it motivates a broader interest in frequency-multiplexed control schemes that share a reduced number of physical lines across many channels~\cite{he2022control,acharya2023multiplexed,forgione2004enhancements,shi2023multiplexed,takeuchi2024microwave,zhao2024multiplexed}. Existing frequency-multiplexed approaches either use narrow channel spacing, with superposed tones generated by room-temperature electronics and separated by on-chip bandpass filters~\cite{shi2023multiplexed}, or rely on large frequency spacing between channels, generated by external high-frequency sources, to maintain adequate isolation~\cite{anferov2024superconducting,anferov2025millimeter}; in both cases the tones originate in room-temperature electronics, and access to the required frequency range in turn demands complex cryogenic filtering and wiring. An on-chip element capable of generating a full comb of well-isolated, high-frequency tones from a single low-frequency input — while providing directional isolation between the generated tones — would address this wiring bottleneck directly at its source, independent of the specific load being driven.

Space-time-periodic modulation has emerged as a powerful route to advanced microwave functionalities, including nonreciprocity, frequency conversion, and parametric amplification, without the bulk and loss penalties of conventional ferrite-based nonreciprocal components~\cite{Taravati_Kishk_MicMag_2019,Taravati_ACSP_2022,taravati2024spatiotemporal,taravati20234d,Taravati_Kishk_PRB_2018,taravati2024efficient,zhang2021magnetic,taravati2025light,zhuang2024superconducting}. Ferrite-free nonreciprocity in superconducting circuits can also be obtained by reservoir engineering, in which a coherent coupling is balanced against an engineered dissipative coupling, yielding on-chip isolators and directional quantum-limited amplifiers~\cite{metelmann2015nonreciprocal}. That approach requires a tuned balance condition between the coherent and dissipative couplings and does not itself generate a harmonic comb, whereas the space-time-periodic bus considered here combines directional propagation with frequency multiplication in a single traveling-wave structure. By simultaneously modulating a transmission line's electromagnetic properties in both space and time, one can break time-reversal symmetry and engineer directional wave propagation, harmonic generation, and gain in a single compact structure~\cite{Taravati_PRAp_2018,taravati2020full,taravati2025_entangle,Taravati_AMA_PRApp_2020,taravati2025finite,taravati2025designing,taravati2021pure,Taravati_PRB_Mixer_2018,taravati_PRApp_2019}. In superconducting circuits, Josephson-junction nonlinearity provides a natural, ultra-low-loss modulation mechanism for realizing such space-time-periodic structures at cryogenic temperatures, making them attractive building blocks for compact, on-chip microwave signal generation and routing, with potential extension to millimeter-wave frequencies~\cite{taravati2024one,taravati2025light,taravati2025_entangle}.

In this paper, we present an on-chip nonreciprocal space-time-periodic Josephson frequency multiplier that performs exactly this function. A single low-frequency input tone at $\omega_\text{m}$ is parametrically up-converted into a comb of high-order harmonics $N\omega_\text{m}$, while the space-time modulation simultaneously enforces nonreciprocal, directional signal propagation along the bus. We develop the space-time-modulated inductance model underlying the device, present full-wave time-domain simulations of the harmonic-generation process under different modulation conditions, and show, in simulation, frequency-selective addressability of individual output channels via the modulation bias point. We further discuss the resulting nonreciprocal isolation between channels and the associated design trade-offs. Because the device delivers a comb of discrete, mutually isolated frequency channels from a single low-frequency source, it is naturally suited to applications requiring multiplexed delivery of many tones to spatially distributed, frequency-addressable loads; we discuss multiplexed qubit control as one such potential application, while noting the additional circuit- and system-level analysis such an application would require beyond the device-level results presented here.

\begin{figure*}
	\begin{center}
		\includegraphics[width=2\columnwidth]{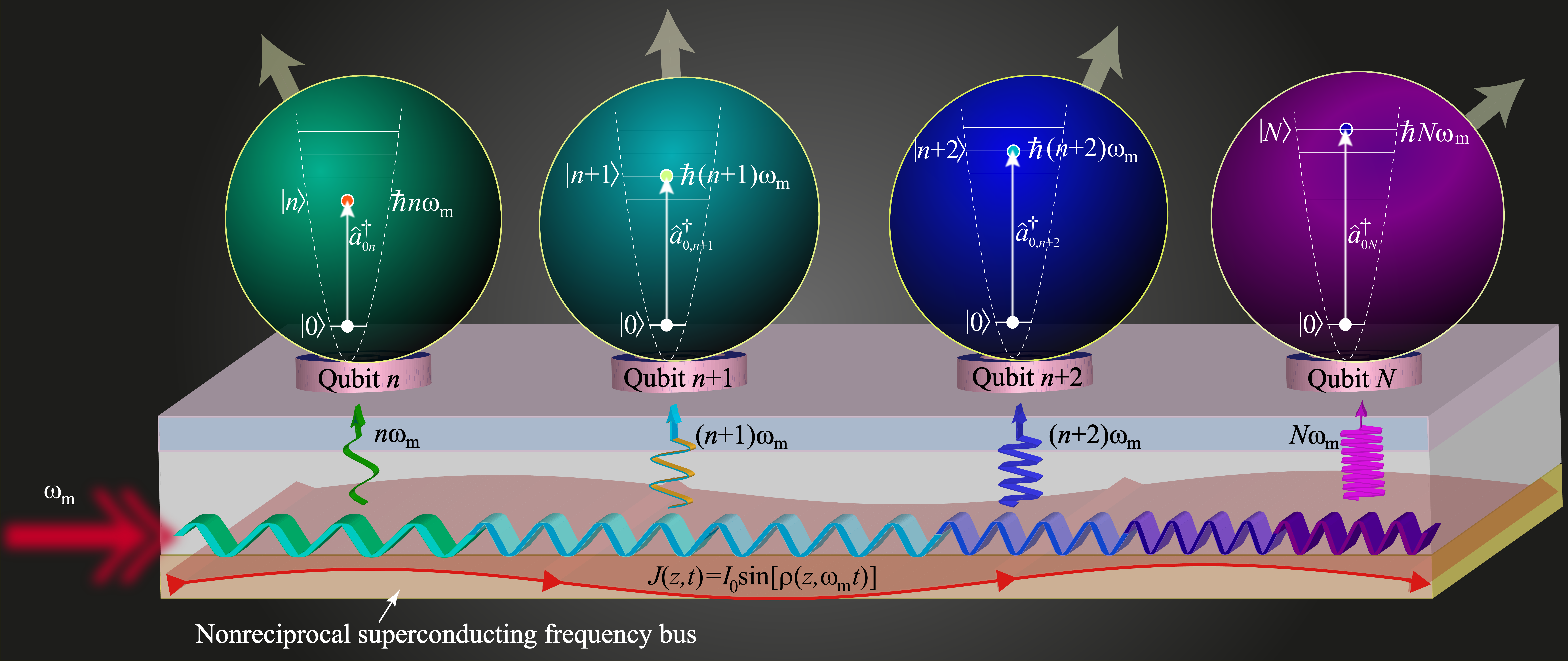}
		\caption{Conceptual architecture of the nonreciprocal frequency‑multiplexed superconducting quantum processor. A space‑time‑periodic superconducting current density \(J(z,t) = I_0 \sin[\rho(z, \omega_\text{m} t)]\) parametrically generates a frequency comb spanning harmonics \(n\omega_\text{m}, (n+1)\omega_\text{m}, \dots, N\omega_\text{m}\). Each harmonic resonantly addresses a distinct flux‑tunable transmon qubit (\(Q_n, Q_{n+1}, \dots, Q_N\)) via its respective transition, enabled by the nonreciprocal superconducting frequency bus. The bus provides directional signal flow (arrows) while suppressing both Purcell decay and inter‑qubit crosstalk, enabling a single low‑frequency input \(\omega_\text{m}\) to control an entire array of qubits with low crosstalk (Sec.~\ref{sec:qsim}).}
		\label{Fig:conc}
	\end{center}
\end{figure*}

\section{Theoretical Implications}\label{sec:theory}
Figure~\ref{Fig:conc} presents the concept of the nonreciprocal space-time-periodic superconducting frequency-multiplier bus as the universal XY drive for a multi-frequency superconducting qubit array. The architecture enables scalable quantum processing by integrating an on-chip space-time-periodic frequency multiplier with a linear array of flux-tunable transmon qubits. A single low-frequency RF input to the multiplier is parametrically up-converted into a harmonic comb, each component of which resonantly addresses a distinct qubit. Individual qubit frequencies are set via dedicated DC flux-bias lines (Z-control), while the nonreciprocal nature of the bus provides directional isolation between the forward drive path and the backward emission path seen by each qubit, as detailed below. This structure is designed to overcome the challenge of wiring complexity and crosstalk in large quantum processors by integrating the control electronics directly beneath the qubit array. The top layer consists of an array of individual Josephson-junction superconducting qubits ($Q_n$ to $Q_N$). Each qubit has a precisely tuned, distinct operating frequency, $\omega_{q_i}$. In the example, these are harmonically related to the fundamental modulation frequency ($\omega_m$): $n\omega_m$, $(n+1)\omega_m$, $(n+2)\omega_m$, and $N\omega_m$. The qubits are arranged spatially, typically with the lowest-frequency qubit ($Q_n$ at $n\omega_m$) placed first and the highest-frequency qubit ($Q_N$ at $N\omega_m$) placed last, relative to the bus's propagation direction. The insets for each qubit show the energy levels ($|0\rangle, |n\rangle, |(n+1)\rangle, \dots$) and the transition frequency $\omega_{q_i}$ that must be addressed for XY control.

\subsection{Hamiltonian Analysis of the Structure}
The entire system Hamiltonian ($\mathcal{H}_{T}$) can be decomposed into three main parts: the Qubit Hamiltonians, the Bus Hamiltonian, and the Interaction Hamiltonians.
\begin{equation}\label{eqa:HT}
	\mathcal{H}_{T} = \sum_i \mathcal{H}_{Q_i} + \mathcal{H}_{B} + \sum_i \mathcal{H}_{I,Q_i-B} + \sum_{i,j} \mathcal{H}_{I,Q_i-Q_j}.
\end{equation}
A dedicated readout resonator $R_i$, dispersively coupled to each qubit in the standard cQED manner~\cite{blais2004cavity,koch2007charge}, provides frequency-multiplexed state readout via a shared feedline (Fig.~\ref{Fig:measd}). Since readout is unaffected by the choice of XY drive architecture, we omit the readout Hamiltonian here and focus on the qubit, bus, and drive-coupling terms that are directly shaped by the proposed nonreciprocal bus.
Qubit Hamiltonian ($\mathcal{H}_{Q_i}$): Each transmon qubit ($Q_i$) is a weakly anharmonic oscillator. Since the qubits are flux-tunable (they have SQUID loops), their Hamiltonian reads
\begin{subequations}\label{eqa:HQi}
	\begin{equation}\label{eqa:HQ}
		\mathcal{H}_{Q_i} = 4 E_{C_i} (\hat{N}_i - N_{g_i})^2 - E_{J_i} \cos(\hat{\phi}_i),
	\end{equation}
	where $E_{C_i}$ is the single-qubit charging energy, and $E_{J_i}$ is the Josephson energy, which is tunable via the SQUID loop flux $\Phi$ (controlled by the Z-line), as
	\begin{equation}\label{eqa:EJ}
		E_{J_i}(\Phi) = 2 E_{J,\text{max}} \cos(\pi \Phi / \Phi_0),
	\end{equation}
	where $N_{g_i}$ is the background charge offset (source of charge noise). In the weakly anharmonic (Duffing) approximation, truncated to the lowest levels, this simplifies to the standard qubit Hamiltonian
	\begin{equation}\label{eqa:HQa}
		\mathcal{H}_{Q_i} \approx \hbar \omega_{q_i} \hat{a}_i^\dagger \hat{a}_i + \frac{\hbar \alpha_i}{2} (\hat{a}_i^\dagger \hat{a}_i)^2,
	\end{equation}
\end{subequations}
where $\omega_{q_i}$ is the qubit frequency and $\alpha_i$ is the anharmonicity.

Bus Hamiltonian ($\mathcal{H}_{B}$): The nonreciprocal bus, acting as a directional, frequency-selective drive line, is a space-time-periodic superconducting frequency multiplier. It is a dynamic nonlinear array whose space-time-varying inductance reads
\begin{subequations}\label{eqa:Lm}
	\begin{equation}\label{eqa:L}
		\begin{split}
			L_\text{m}(I,z,t)
			= \dfrac{\Phi_0}{2\pi I_0}\sec\left(\widetilde{\Phi}_\text{dc}+ \widetilde{\Phi}_\text{rf} \sin[\kappa_\text{s} z-\omega_\text{s} t+\phi]\right),
		\end{split}
	\end{equation}
	where $\widetilde{\Phi}_\text{dc}= 2\pi \Phi_\text{dc}/ \Phi_0$ and $\widetilde{\Phi}_\text{rf}= 2\pi \Phi_\text{rf}/ \Phi_0$, and $\Phi_0$ is the magnetic flux quantum. The tilded quantities are dimensionless phases in radians, and Eq.~\eqref{eqa:L} is finite only while $\widetilde{\Phi}_\text{dc}+\widetilde{\Phi}_\text{rf}<\pi/2$, since the secant diverges as the junction phase approaches $\pi/2$. The Hamiltonian of the bus reads
	\begin{equation}\label{eqa:HB}
		\mathcal{H}_{B} = \sum_k \hbar \omega_k \hat{b}_k^\dagger \hat{b}_k + \mathcal{H}_{NL}(t),
	\end{equation}
\end{subequations}
where $\mathcal{H}_{NL}(t)$ contains the nonlinear interaction terms generated by the Josephson junctions, modulated in space and time by the input current $J(z,t)$ at $\omega_m$. This nonlinearity is the source of both the frequency multiplication (creating harmonics $N\omega_m$) and the nonreciprocity: because the modulation pattern in Eq.~\eqref{eqa:L} propagates unidirectionally (left to right, at $\kappa_\text{s}, \omega_\text{s}$), the bus's mode structure at each harmonic frequency is not symmetric between the forward ($+z$) and backward ($-z$) propagation directions. This asymmetry is the physical origin of the nonreciprocal coupling introduced below.

Qubit-bus interaction Hamiltonian ($\mathcal{H}_{I,Q_i-B}$): Because the bus supports distinct forward- and backward-propagating mode structures at each harmonic, a qubit $Q_i$ coupled to the bus near position $z_i$ couples to two counter-propagating bus modes at its operating frequency $\omega_{q_i} = N_i\omega_m$: a forward mode $\hat{b}_{\omega_{q_i}}^{(+)}$, carrying the drive signal from the input port toward the qubit array, and a backward mode $\hat{b}_{\omega_{q_i}}^{(-)}$, propagating from the qubit array back toward the input port. The interaction Hamiltonian is
\begin{equation}\label{eqa:HQB}
	\begin{split}
		\mathcal{H}_{I,Q_i-B} \approx \; & \hbar g_i^{(+)} \left(\hat{a}_i^\dagger \hat{b}_{\omega_{q_i}}^{(+)} e^{-iN_i\omega_m t} + \text{h.c.}\right) \\
		& + \hbar g_i^{(-)} \left(\hat{a}_i^\dagger \hat{b}_{\omega_{q_i}}^{(-)} e^{-iN_i\omega_m t} + \text{h.c.}\right),
	\end{split}
\end{equation}
where $g_i^{(+)}$ and $g_i^{(-)}$ are the qubit's coupling strengths to the forward and backward bus modes, respectively, and $N_i\omega_m$ is the harmonic frequency generated by the bus that drives $Q_i$ (e.g., $n\omega_m$ for $Q_1$, $N\omega_m$ for $Q_4$). In a conventional reciprocal bus, $g_i^{(+)} = g_i^{(-)}$ by symmetry. In the proposed space-time-modulated bus, the unidirectional propagation of the modulation pattern breaks this symmetry, so that $g_i^{(-)} \ll g_i^{(+)}$: the backward mode at $\omega_{q_i}$ is only weakly populated relative to the forward-launched drive. We define the directional asymmetry ratio
\begin{equation}\label{eqa:kappa}
	\kappa_i \equiv \frac{g_i^{(-)}}{g_i^{(+)}} \ll 1,
\end{equation}
which quantifies the degree of nonreciprocal isolation seen by qubit $Q_i$, and which can in principle be extracted directly from the bus's simulated or measured forward/backward transmission asymmetry at $\omega_{q_i}$ (e.g., $|S_{21}|$ vs.\ $|S_{12}|$), rather than assumed a priori. This directional coupling picture parallels the theory of cascaded quantum systems, in which one mode drives another without back-action~\cite{metelmann2015nonreciprocal}; here the asymmetry is set by the traveling-wave modulation, and $\kappa_i=0$ would correspond to the ideal cascaded limit.

Qubit-Qubit Interaction Hamiltonian ($\mathcal{H}_{I,Q_i-Q_j}$): This term describes the unwanted direct coupling, or static crosstalk, between neighboring qubits ($Q_i$ and $Q_j$), which is a major source of error in multi-qubit systems. In planar superconducting circuits, qubits are placed close to each other, and the dominant coupling mechanism is typically residual capacitive or inductive coupling through the chip substrate and ground plane. In the dispersive regime, the interaction Hamiltonian between two qubits simplifies to a $\hat{\sigma}_x\hat{\sigma}_x$ (or $\hat{\sigma}_z\hat{\sigma}_z$) interaction, often approximated as an Ising-type interaction, but the primary error mechanism is the $\hat{\sigma}_x\hat{\sigma}_x$ term
\begin{equation}\label{eqa:HQQ}
	\mathcal{H}_{I,Q_i-Q_j} = g_{ij}(\hat{\sigma}_i^x\hat{\sigma}_j^x + \hat{\sigma}_i^y\hat{\sigma}_j^y),
\end{equation}
where $g_{ij}$ is the direct coupling strength between $Q_i$ and $Q_j$. This strength must be minimized by design (e.g., maximizing the distance between qubits) to prevent unwanted two-qubit gates and coherent crosstalk during single-qubit operations. The use of a frequency-multiplexed bus with widely separated qubit frequencies is a major advantage here. Since the qubit frequencies are detuned by $\Delta_{ij}$, the static $\mathcal{H}_{I,Q_i-Q_j}$ interaction is highly off-resonant. The interaction primarily causes frequency pull (a small static shift in each other's frequency) rather than fast, unwanted transitions, allowing $Q_i$ and $Q_j$ to be treated largely independently during single-qubit operations.

\subsection{Additional Noise Considerations}
Beyond the coupling-driven Purcell and crosstalk effects analyzed below, two additional noise sources affect qubit coherence in this architecture. First, the strong pump tone ($\omega_m$) used to modulate the bus is itself a potential noise source that can reduce $T_2$; careful filtering and noise suppression of the $\omega_m$ pump is therefore critical to maintaining high coherence. Second, since the qubits are flux-tunable via SQUID loops, they are susceptible to low-frequency magnetic flux noise (Z-noise), which limits $T_2^*$, and the small SQUID loops required at high operating frequencies can increase sensitivity to this noise.

\subsubsection{Purcell Decay Analysis ($\Gamma_{\text{Purcell}}$)}
Purcell decay is the enhancement of the spontaneous emission rate ($\Gamma_1 = 1/T_1$) of the qubit due to its coupling to the electromagnetic environment of the bus. Using the directional coupling picture of Eq.~\eqref{eqa:HQB}, an excited, undriven qubit radiates into both bus modes, so the total bus-induced decay rate is proportional to $(g_i^{(+)})^2+(g_i^{(-)})^2$. The share emitted backward, toward the shared input line, is the channel that can return noise and loss to the other qubits, and it scales as:
\begin{equation}\label{eqa:GammaPurcellBus}
	\Gamma_{\text{Purcell,Bus}}^{(-)}(\omega_{q_i}) \propto \left(g_i^{(-)}\right)^2 = \kappa_i^2 \left(g_i^{(+)}\right)^2,
\end{equation}
where $\kappa_i$ is the directional asymmetry ratio of Eq.~\eqref{eqa:kappa}. For a reciprocal bus, $\kappa_i = 1$ and the qubit decays symmetrically in both directions at a rate set by $g_i^{(+)}$ alone. For the proposed nonreciprocal bus, $\kappa_i \ll 1$, so the backward-coupled Purcell channel is suppressed by a factor $\kappa_i^2$ relative to a reciprocal bus of the same forward coupling strength. The forward channel is not suppressed: the total bus-induced rate is $\Gamma_{\text{Purcell,Bus}}\propto(1+\kappa_i^2)(g_i^{(+)})^2$, which is at most a factor of two below that of a reciprocal bus with the same $g_i^{(+)}$, and its magnitude depends on how the forward port is terminated. The benefit of nonreciprocity is therefore the redirection of emission away from the shared input line, not a large reduction of the total decay rate. This provides a coupling-level, rather than purely impedance-level, picture of Purcell suppression: the qubit's emission is not eliminated but is redirected predominantly into the forward-propagating mode, which is routed away from the qubit toward the rest of the array or an output termination rather than back through a shared input line. Quantifying $\Gamma_{\text{Purcell,Bus}}$ for a given qubit therefore reduces to characterizing $\kappa_i$ from the bus's simulated or measured directional transmission at $\omega_{q_i}$, which can be obtained from the same full-wave simulation framework used in Sec.~\ref{sec:res}, rather than requiring an independent admittance calculation. It is important to note that while the proposed nonreciprocal bus can suppress this backward-coupled Purcell channel, in state-of-the-art superconducting qubits the dominant sources of decoherence are typically intrinsic material losses, two-level system (TLS) defects, and quasiparticle generation, particularly at millimeter-wave frequencies. Operating qubits at higher frequencies (e.g., 20--75\,GHz) generally increases sensitivity to surface dielectric losses and quasiparticle poisoning, which may offset some of the benefits of reduced Purcell decay. The architecture presented here is therefore primarily intended to address the wiring complexity and crosstalk bottlenecks, while offering a pathway toward improved coherence as materials and fabrication techniques mature.

\subsubsection{Crosstalk Analysis ($\mathcal{C}_{ij}$)}
Crosstalk refers to the unwanted coupling between two qubits, $Q_i$ and $Q_j$, when a control pulse intended for $Q_i$ unintentionally affects $Q_j$. In the frequency-multiplexed system, the primary concern is coherent crosstalk through the shared bus. If a drive tone at $\omega_{q_i}$ is injected, the effective Hamiltonian seen by $Q_j$ is
\begin{equation}\label{eqa:Hcrosstalk}
	\mathcal{H}_{\text{Crosstalk}} = \hbar\Omega_{\text{eff},j}\left(\hat{\sigma}_j^+ e^{-i\omega_{q_i}t} + \hat{\sigma}_j^- e^{i\omega_{q_i}t}\right),
\end{equation}
where $\Omega_{\text{eff},j}$ is the effective Rabi frequency induced on $Q_j$ by the drive intended for $Q_i$. The architecture suppresses this crosstalk through two complementary mechanisms.

\begin{itemize}
	\item Frequency Detuning (Passive Suppression): The qubits are widely spaced in frequency, which leads to inherently small crosstalk. The coherent crosstalk coefficient ($\mathcal{C}_{ij}$) for $Q_j$ due to the drive of $Q_i$ scales inversely with the detuning ($\Delta_{ij} = |\omega_{q_i}-\omega_{q_j}|$) when the tones propagate through a linear medium, as
	\begin{equation}\label{eqa:Cij}
		\mathcal{C}_{ij} \propto \frac{1}{|\omega_{q_i}-\omega_{q_j}|}.
	\end{equation}
	
	\item Nonreciprocal Bus Isolation (Active Suppression): In addition to frequency detuning, the directional asymmetry of the bus provides an active form of isolation. Because a drive tone launched from the input port predominantly excites the forward mode at each qubit's location, off-resonant leakage to a distant qubit $Q_j$ must propagate through the same forward-favored channel, and any component that would otherwise reflect and scatter backward toward $Q_i$ is suppressed by the same $\kappa$ factor introduced in Eq.~\eqref{eqa:kappa}. This reduces reflections and back-propagation that could otherwise create unwanted coupling paths between distant qubits, enhancing the isolation already provided by frequency detuning.
\end{itemize}

\subsection{Theoretical Framework and Error Model}
The performance of the qubit array can be assessed using the total single-qubit gate error rate, $E_{\text{gate}}$, modeled as the sum of a non-unitary coherence error and a unitary coherent-crosstalk error,
\begin{subequations}
	\begin{equation}\label{eqa:Egate}
		E_{\text{gate}}(\omega_{q_i}) = E_{\text{Coherence}}(\omega_{q_i}) + E_{\text{Crosstalk}}(\omega_{q_i},N).
	\end{equation}
	This model incorporates the frequency-dependent parameters of the $i$-th qubit and the array size $N$. The coherence error, $E_{\text{Coherence}}$, arises from energy relaxation during the fixed gate duration $T_{\text{gate}}$:
	\begin{equation}\label{eqa:Ecoh}
		E_{\text{Coherence}}(\omega_{q_i}) = T_{\text{gate}} \cdot \Gamma_1^{\text{total}}(\omega_{q_i}),
	\end{equation}
	where the total relaxation rate combines intrinsic material losses and the total (forward plus backward) bus-induced Purcell rate:
	\begin{equation}\label{eqa:Gamma1total}
		\Gamma_1^{\text{total}}(\omega_{q_i}) = \Gamma_1^{\text{intrinsic}} + \Gamma_{\text{Purcell,Bus}}(\omega_{q_i}).
	\end{equation}
	The coherent crosstalk error, $E_{\text{Crosstalk}}$, quantifies the unwanted rotation induced on non-target qubits by the drive signal intended for $Q_i$, and is the primary mechanism limiting scalability in a shared-bus architecture:
	\begin{equation}\label{eqa:Ecrosstalk}
		E_{\text{Crosstalk}}(\omega_{q_i},N) = \sum_{j\neq i}^{N} \left(\frac{g_{\text{eff}}}{\Delta\omega_{ij}}\right)^2 \kappa_{ij},
	\end{equation}
\end{subequations}
which is a dimensionless excitation probability: for a smooth pulse of duration $T_{\text{gate}}\gg 1/\Delta\omega_{ij}$, the off-resonant excitation is set by the ratio $(g_{\text{eff}}/\Delta\omega_{ij})^2$ rather than growing with $T_{\text{gate}}$ (see Sec.~\ref{sec:qsim}). Here $g_{\text{eff}}$ is the effective qubit-to-qubit coupling strength via the bus, $\Delta\omega_{ij} = |\omega_{q_i}-\omega_{q_j}|$ is the frequency separation between qubits (a multiple of $\omega_m$), and $\kappa_{ij}$ is the directional isolation factor between the drive channel of $Q_i$ and the location of $Q_j$, generalizing the single-qubit asymmetry ratio $\kappa_i$ of Eq.~\eqref{eqa:kappa} to a qubit pair. Because $\Gamma_{\text{Purcell,Bus}}$ and $\kappa_{ij}$ are both defined directly in terms of the bus's directional coupling and transmission characteristics, evaluating Eqs.~\eqref{eqa:Gamma1total} and~\eqref{eqa:Ecrosstalk} for a specific design requires characterizing these quantities from full-wave simulation or measurement of the bus, rather than from an assumed impedance model.

\section{Results}\label{sec:res} 
\subsection{Integration of the frequency multiplier control bus with distinct-frequency qubits}
\begin{figure*}
	\begin{center}
		\includegraphics[width=1.8\columnwidth]{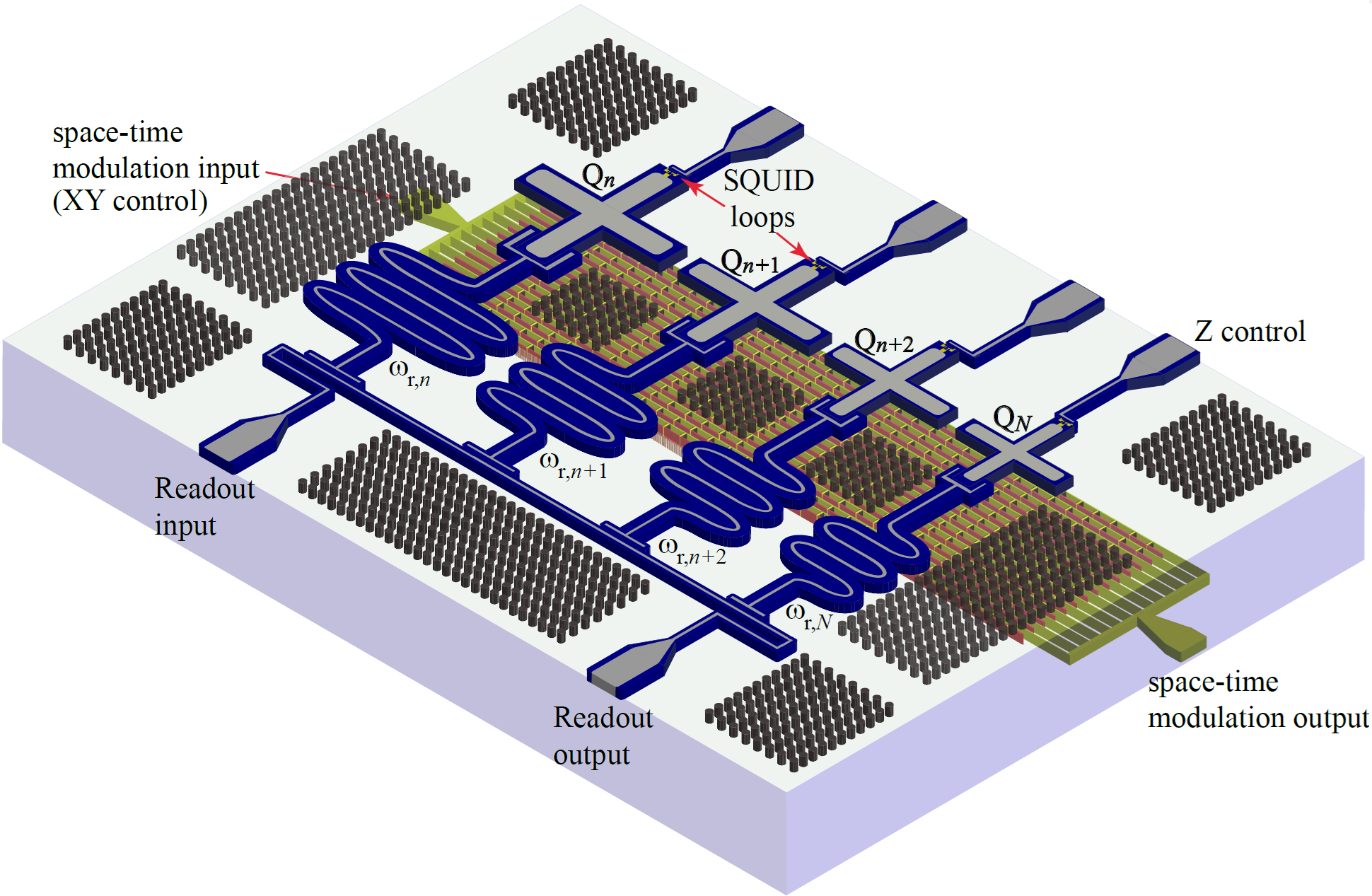}
		\caption{ Integrated on-chip architecture for frequency-multiplexed superconducting qubits. The device consists of a linear array of flux-tunable transmon qubits (top) coupled to a space‑time‑modulated superconducting frequency multiplier that serves as a nonreciprocal control bus (bottom). Each qubit's SQUID loop is addressed by a dedicated DC flux bias (Z‑control), setting its frequency to a distinct harmonic of the modulation frequency. A single low‑frequency RF input drives the parametric comb generation in the multiplier, while the nonreciprocal bus provides directional isolation to suppress Purcell decay and crosstalk, supporting scalable multiplexed operation.}
		\label{Fig:sch}
	\end{center}
\end{figure*}

Figure~\ref{Fig:sch} illustrates the integrated space-time-periodic frequency multiplier control bus with an array of distinct-frequency qubits. The nonreciprocal spatiotemporal frequency bus in the bottom layer is the core innovation, serving as the multiplexed XY control bus: an array of Josephson junctions forming a superconducting structure whose properties are periodic in both space ($z$) and time ($t$)~\cite{taravati2024spatiotemporal,taravati2025light,taravati2025_entangle}. A low-frequency input signal at $\omega_m$ is injected into the multiplier (large red arrow); the nonlinearity of the Josephson junctions, combined with the space-time modulation, up-converts this input into a comb of high-order harmonics $N\omega_m$. The unidirectional propagation of the modulation pattern (red arrow on the current $J(z,t)$) makes this generation process inherently nonreciprocal, so the harmonic comb propagates predominantly forward along the bus rather than symmetrically in both directions — the directional coupling asymmetry underlying the isolation discussed in Sec.~\ref{sec:theory}. Due to precise frequency matching, each qubit couples resonantly only to the specific harmonic $N_i\omega_m$ generated at its location: for example, the $n\omega_m$ harmonic couples strongly to $Q_n$ alone. This allows a single, centralized multiplier to deliver all necessary XY control signals to the entire qubit array from one low-frequency source, substantially reducing the wiring complexity of the cryogenic setup relative to an architecture requiring a dedicated high-frequency line per qubit.

\begin{figure}
	\begin{center}
		\includegraphics[width=1\columnwidth]{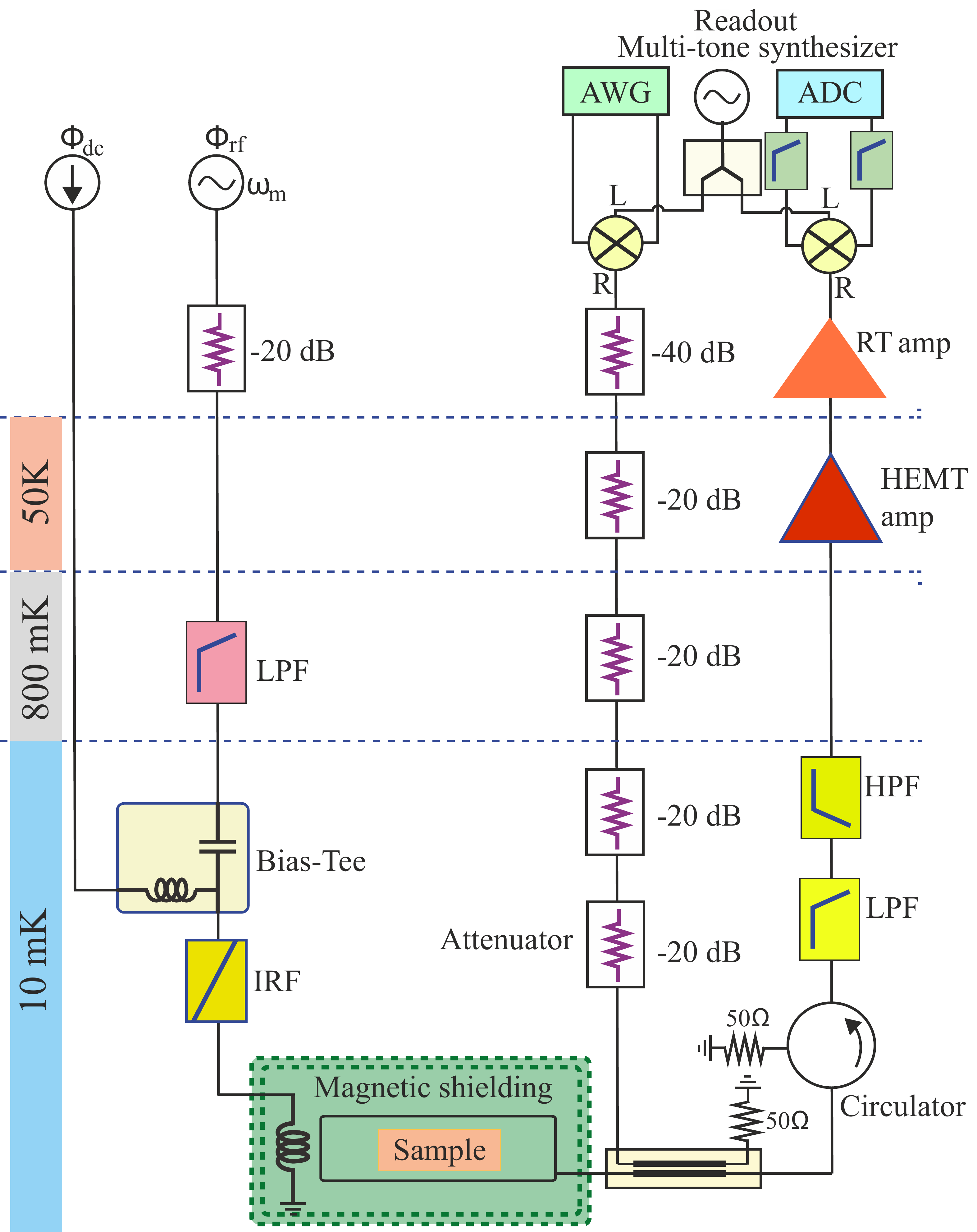}
		\caption{Proposed cryogenic measurement setup for future experimental characterization of the frequency-multiplexed qubit array with the on-chip nonreciprocal space-time-modulated frequency multiplier bus.}
		\label{Fig:measd}
	\end{center}
\end{figure}

\subsection{Cryogenic Measurement Setup}
Figure~\ref{Fig:measd} outlines a proposed cryogenic measurement setup for future experimental characterization of the device, divided into three subsystems — XY control, Z (flux) control, and multiplexed readout — each intended for operation at millikelvin temperatures. The XY control path delivers the low-frequency RF pump tone at $\omega_\text{m}$ to the on-chip Josephson frequency multiplier, attenuated by $-20$\,dB at each of three stages (room temperature, 50\,K, and 800\,mK) to thermalize the line and suppress blackbody radiation, followed by a low-pass filter at 800\,mK with cutoff above $\omega_\text{m}$ (e.g., $2\pi\times4$\,GHz for a $2\pi\times3$\,GHz pump) to allow efficient drive while suppressing noise and higher harmonics. The DC flux bias ($\Phi_{\rm dc}$) line, attenuated by $-20$\,dB at room temperature and 50\,K, combines with the filtered pump tone at a bias-tee at the 10\,mK stage before reaching the sample through an infrared filter (IRF). Independent DC Z-control lines (not shown) would provide individual flux bias to each qubit's SQUID loop for frequency tuning, incorporating similar low-pass filtering to suppress wideband noise that would otherwise dephase the qubits; careful electromagnetic design would be needed to minimize unwanted mutual inductance between these lines and the bus.

Readout would employ frequency-division multiplexing, with a multi-tone synthesizer generating probe signals at each qubit's readout-resonator frequency. These tones, attenuated by $-20$\,dB at room temperature, 50\,K, and 800\,mK, would be combined and sent to the sample through a weakly coupled port via a cryogenic circulator. Reflected signals would pass through high- and low-pass filters, be amplified by a HEMT amplifier at 50\,K and a room-temperature amplifier, and be digitized by an ADC. Magnetic shielding would surround the sample at the 10\,mK stage to minimize flux noise, and superconducting NbTi coaxial cables would be used throughout the cryostat to reduce thermal conductivity and signal loss. Attenuation, filtering, and thermalization stages follow standard practice for millikelvin microwave measurements~\cite{krinner2019engineering}; representative values are shown, though the specific budget has not yet been experimentally optimized for this architecture.

\begin{figure}
	\begin{center}
		\subfigure[]{\label{Fig:dcp6_rf_p6a}
			\includegraphics[width=1\columnwidth]{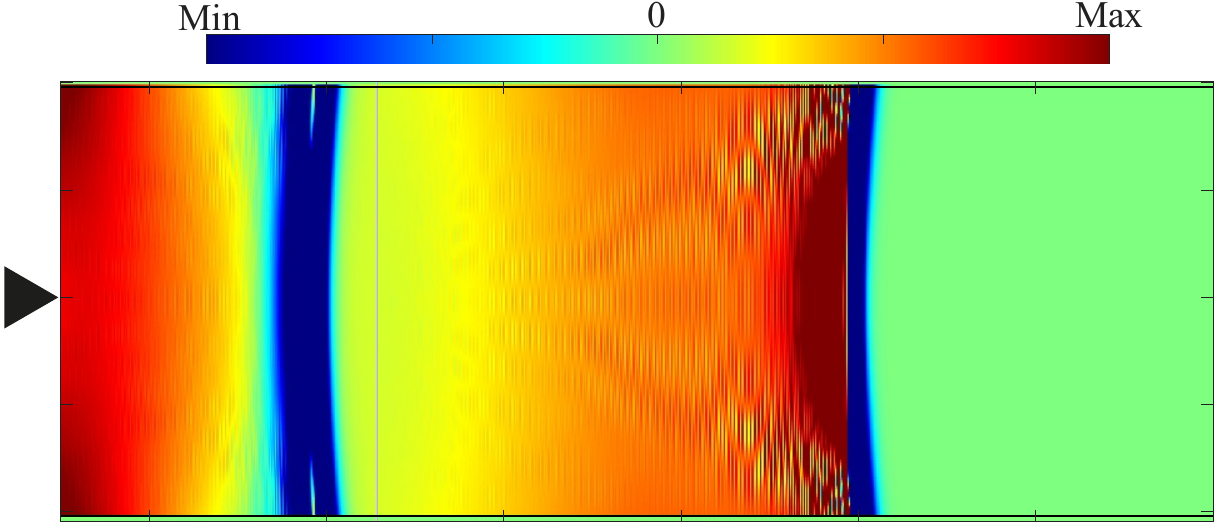}}
		\subfigure[]{\label{Fig:dcp6_rf_p85b}
			\includegraphics[width=1\columnwidth]{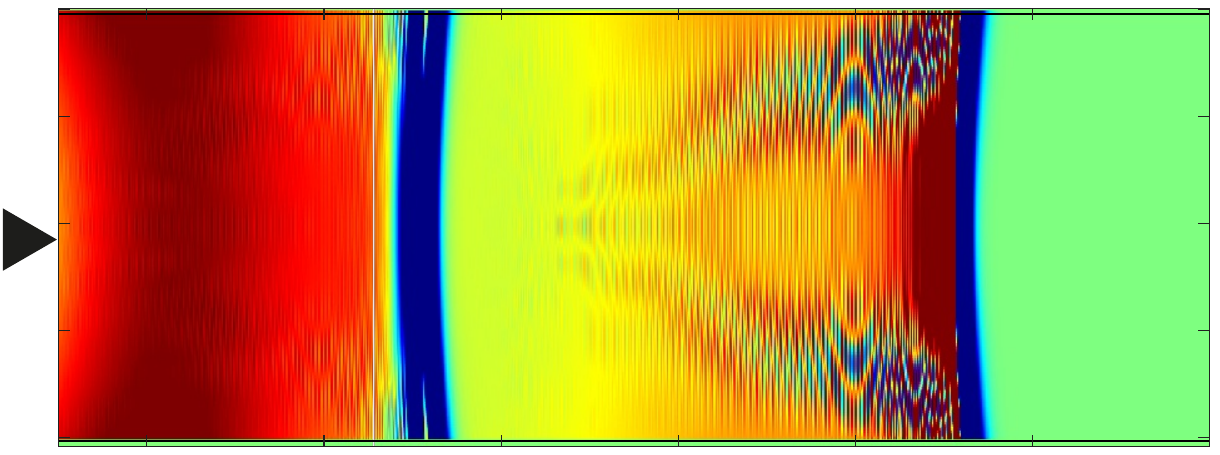}}
		\subfigure[]{\label{Fig:dcp6_rf_p6c}
		\includegraphics[width=1\columnwidth]{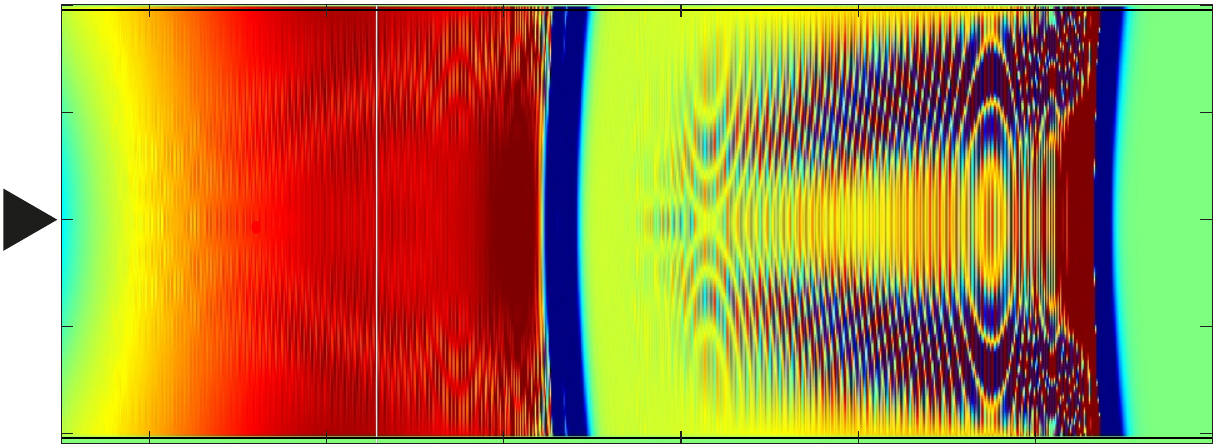}}
		\subfigure[]{\label{Fig:dcp6_rf_p6d}
			\includegraphics[width=1\columnwidth]{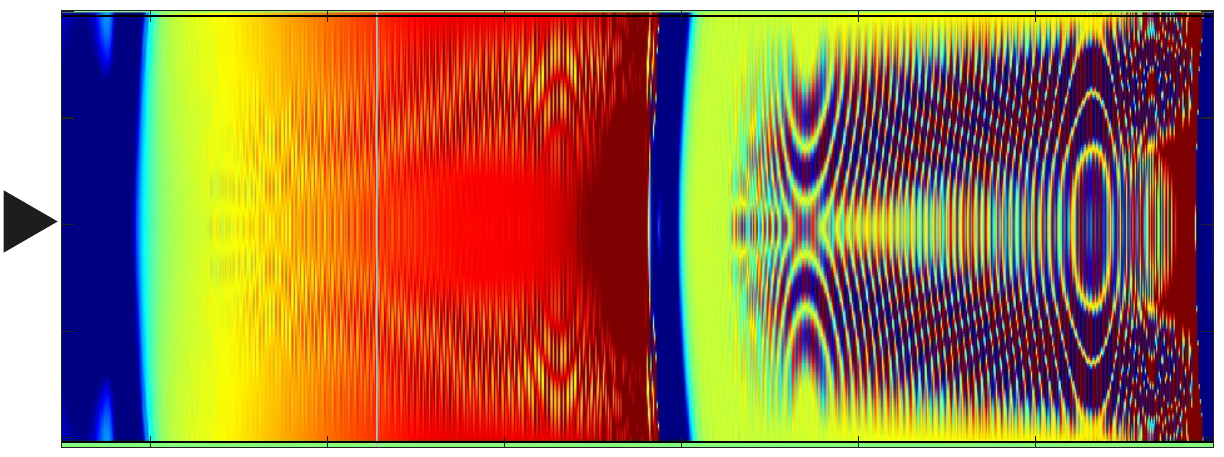}}	
			\subfigure[]{\label{Fig:dcp6_rf_p6e}
		\includegraphics[width=1\columnwidth]{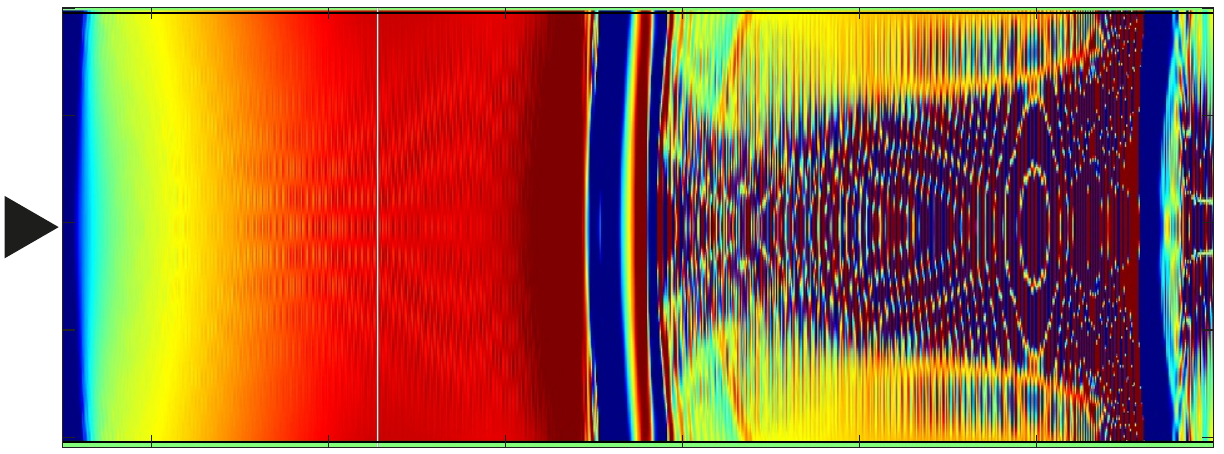}}	
		\caption{Magnetic field distribution for magnetic flux excitation at frequency $\omega_\text{m}$ incident from the left, showing significant frequency multiplication in the spatiotemporal superconducting array. The array parameters are $\widetilde{\Phi}\text{dc}=0.6$ and $\widetilde{\Phi}\text{rf}=0.6$. These time-domain results show the field distribution at successive time steps: (a) $t=0.3$ ns, (b) $t=0.37$ ns, (c) $t=0.5$ ns, (d) $t=0.6$ ns, and (e) $t=0.8$ ns.}
		\label{Fig:dcp6_rf_p6}
	\end{center}
\end{figure}

\subsection{Numerical Simulation Results}
Figures~\ref{Fig:dcp6_rf_p6a}--\ref{Fig:dcp6_rf_p6e} present a time-resolved numerical simulation of the magnetic field distribution within the spatiotemporally modulated array, demonstrating its operation as an integrated frequency multiplier. An excitation at $\omega_\text{m}$ is incident from the left, propagating through an array whose effective Josephson inductance is dynamically modulated by a static flux bias $\widetilde{\Phi}_{\rm dc}=0.6$ and an RF flux drive $\widetilde{\Phi}_{\rm rf}=0.6$. These values were chosen from the parameter sweeps presented below: this combination provides strong parametric frequency conversion while avoiding excessive nonlinear distortion and back-scattering, and is feasible in cryogenic experiments since the modulation is applied globally to the bus rather than directly to individual qubits; potential RF-pump heating effects are expected to be manageable with standard cryogenic attenuation and filtering. The panel sequence, spanning $t=0.3$--$0.8$\,ns, captures the nonlinear wave-mixing process in the time domain: the magnetic field profile evolves from predominantly sinusoidal in (a) to progressively distorted and rich in higher spatial harmonics in (c)--(e), visually manifesting the frequency-multiplication process as the time-periodic modulation breaks time-reversal symmetry and parametrically generates integer harmonics $n\omega_\text{m}$. The increasing spatial oscillation frequency across the array at later time steps (d)--(e) corresponds directly to the generation and propagation of the second, third, and higher harmonics. These simulations indicate that the proposed superconducting metasurface functions as a compact, cryogenic frequency multiplier capable of up-converting a single microwave input into a comb of higher-frequency tones suitable for addressing multi-frequency qubit arrays.

We note that the nonlinear effective refractive index of the modulated array in Eq.~\eqref{eqa:L} produces an amplitude-dependent space-time response in the propagating field — a nonuniform spatial dilation and self-induced spectral redshift qualitatively reminiscent of wave propagation in analogue expanding-spacetime systems~\cite{GamowUniverse,ginis2010frequency,westerberg2014experimental}. We mention this parallel only as a physical interpretation of the observed waveform broadening; a rigorous analogue-gravity treatment is beyond the scope of this device-oriented study.

\begin{figure}
	\begin{center}
		\subfigure[]{\label{Fig:dcp4_rf_p85a}
			\includegraphics[width=1\columnwidth]{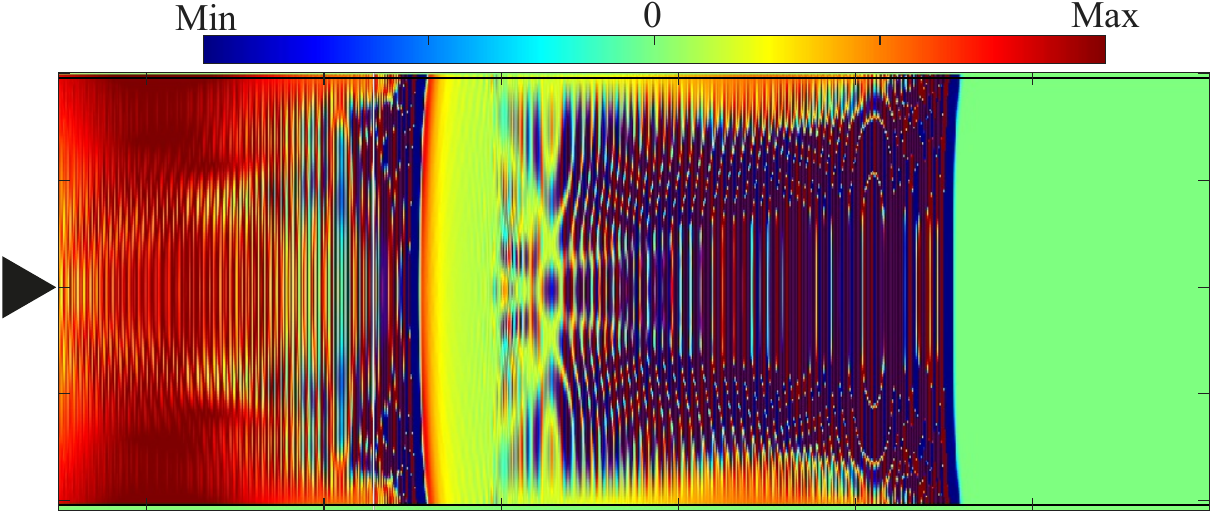}}
		\subfigure[]{\label{Fig:dcp4_rf_p85b}
			\includegraphics[width=1\columnwidth]{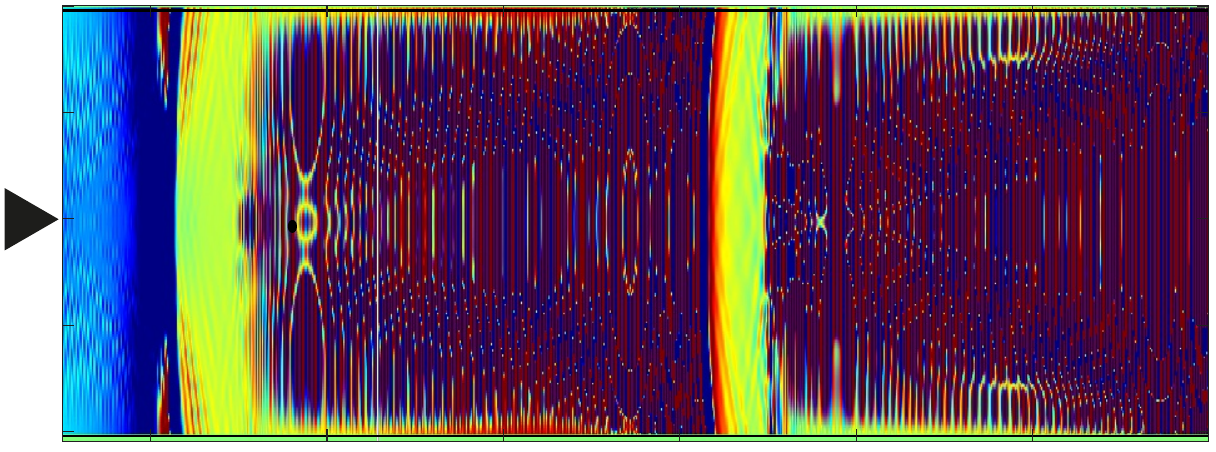}}
		\subfigure[]{\label{Fig:dcp4_rf_p85c}
		\includegraphics[width=1\columnwidth]{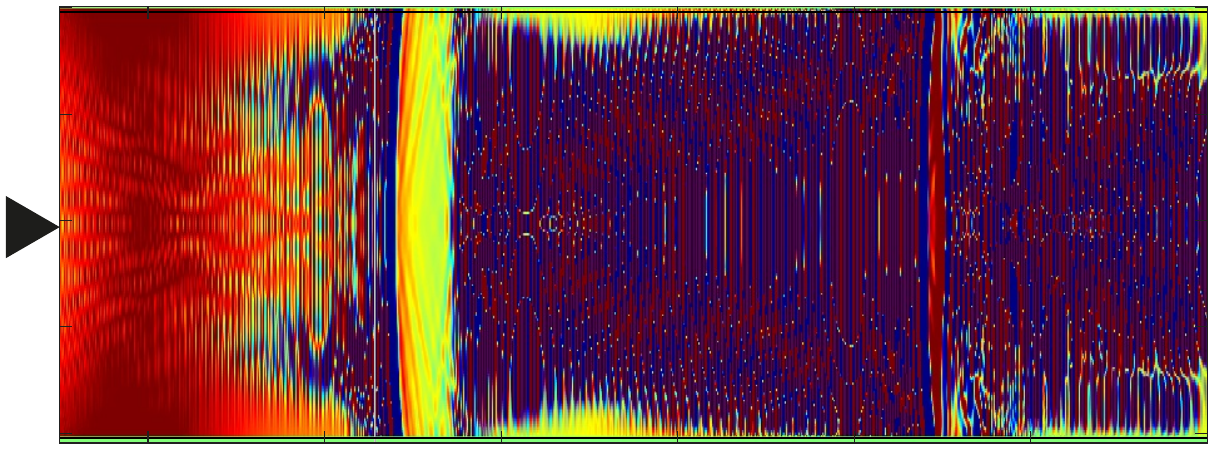}}
	\subfigure[]{\label{Fig:dcp4_rf_p85d}
		\includegraphics[width=1\columnwidth]{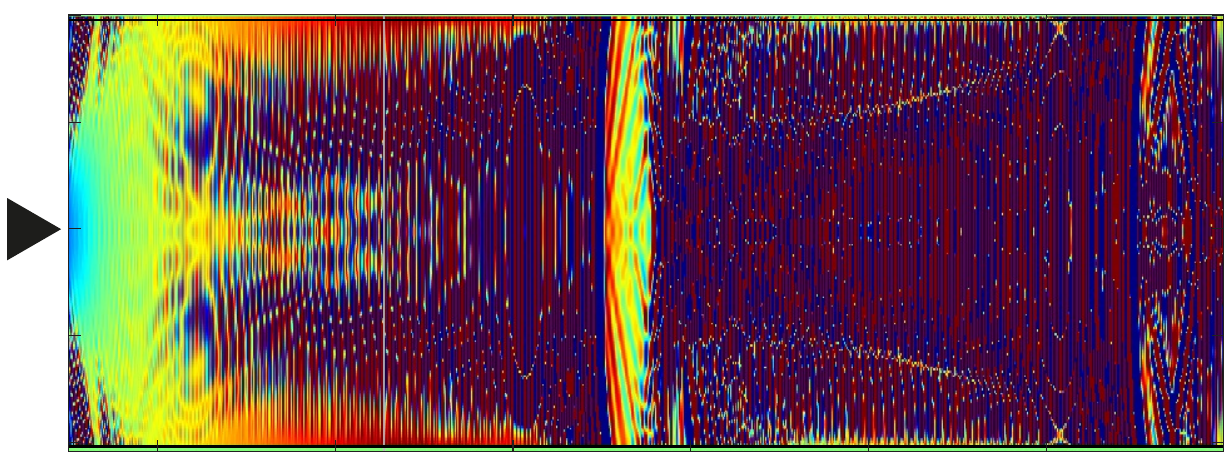}}	
		\caption{Magnetic field distribution for magnetic flux excitation at frequency $\omega_\text{m}$ incident from the left, showing significant frequency multiplication as well as amplification in the spatiotemporal superconducting array. The array parameters are $\widetilde{\Phi}\text{dc}=0.4$ and $\widetilde{\Phi}\text{rf}=0.85$. These time-domain results show the field distribution at successive time steps: (a)~$t=0.4$ ns, (b)~$t=0.55$ ns, (c)~$t=0.7$ ns. (d)~$t=0.8$ ns.}
		\label{Fig:dcp4_rf_p85}
	\end{center}
\end{figure}

Figures~\ref{Fig:dcp4_rf_p85a} to~\ref{Fig:dcp4_rf_p85d} examine the frequency multiplication dynamics under a significantly different flux configuration: a reduced static bias $\widetilde{\Phi}_{\rm dc}=0.4$ paired with a strong RF modulation amplitude $\widetilde{\Phi}_{\rm rf}=0.85$. This parameter set explores a regime where the dynamic flux swing constitutes a large fraction of a flux quantum, driving the Josephson junctions through deep nonlinear excursions. The field pattern develops a highly nonlinear, almost shock-wave-like profile characterized by steep field gradients and compressed wavefronts, indicating vigorous generation of a broad spectrum of higher harmonics. The complex, multi-frequency spatial interference pattern visible in Figs.~\ref{Fig:dcp4_rf_p85c} and~\ref{Fig:dcp4_rf_p85d} suggests the simultaneous excitation of several harmonic orders. This configuration demonstrates the structure's capability for broadband harmonic generation, a crucial feature for applications requiring multi-frequency outputs; however, the stronger spatial distortion relative to the balanced case (Fig.~\ref{Fig:dcp6_rf_p6}) may also indicate increased back-reflection or nonlinear losses, presenting a trade-off between multiplication bandwidth and power transmission. This result highlights the design flexibility of the platform, where the flux parameters can be tuned to prioritize either high harmonic selectivity (Fig.~\ref{Fig:dcp6_rf_p6}) or wideband spectral generation (as shown here).

\begin{figure}
	\begin{center}
		\subfigure[]{\label{Fig:dcp8_rf_p2}
    	\includegraphics[width=1\columnwidth]{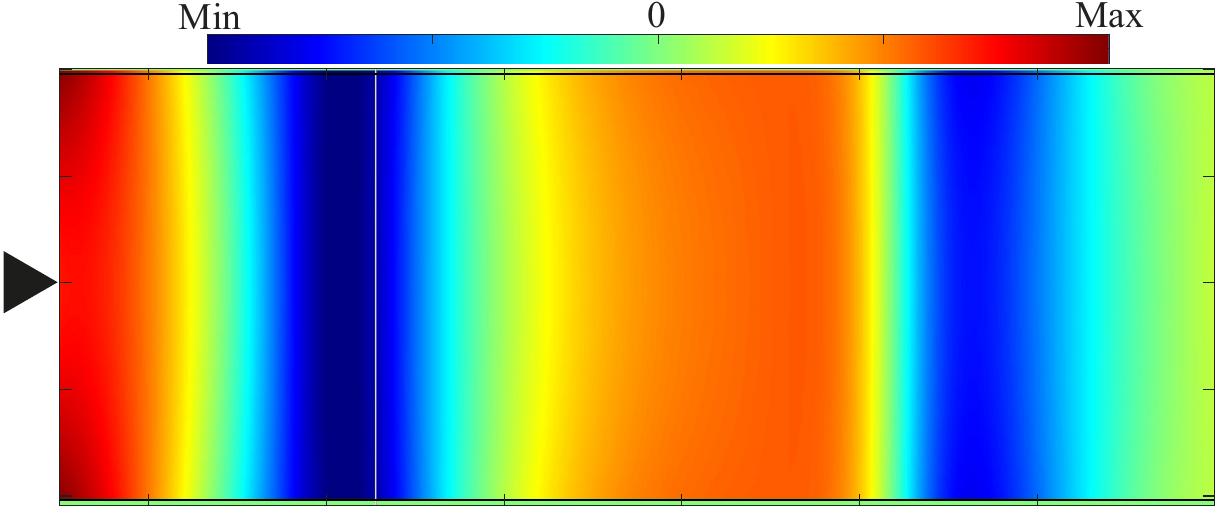}}
		\subfigure[]{\label{Fig:dcp8_rf_p4a}
			\includegraphics[width=1\columnwidth]{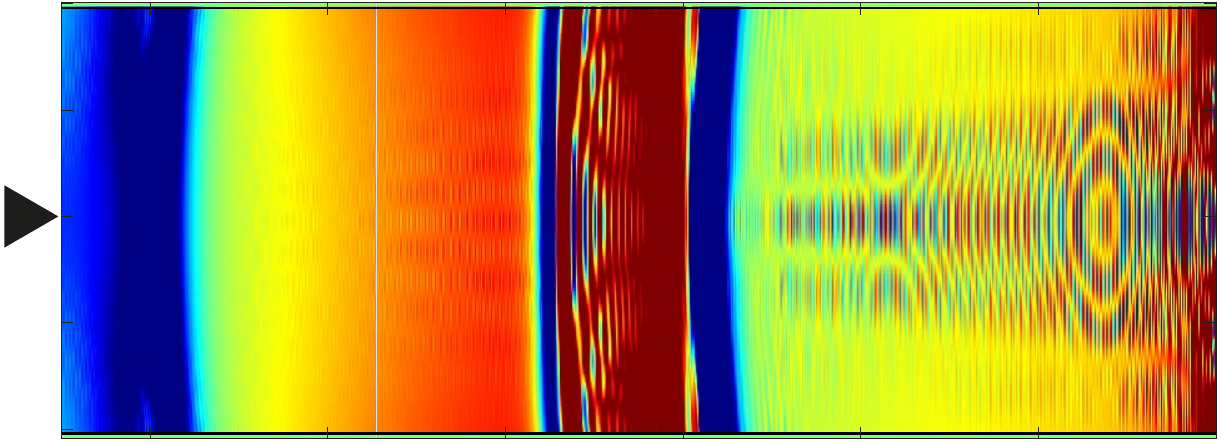}}
		\subfigure[]{\label{Fig:dcp8_rf_p4b}
			\includegraphics[width=1\columnwidth]{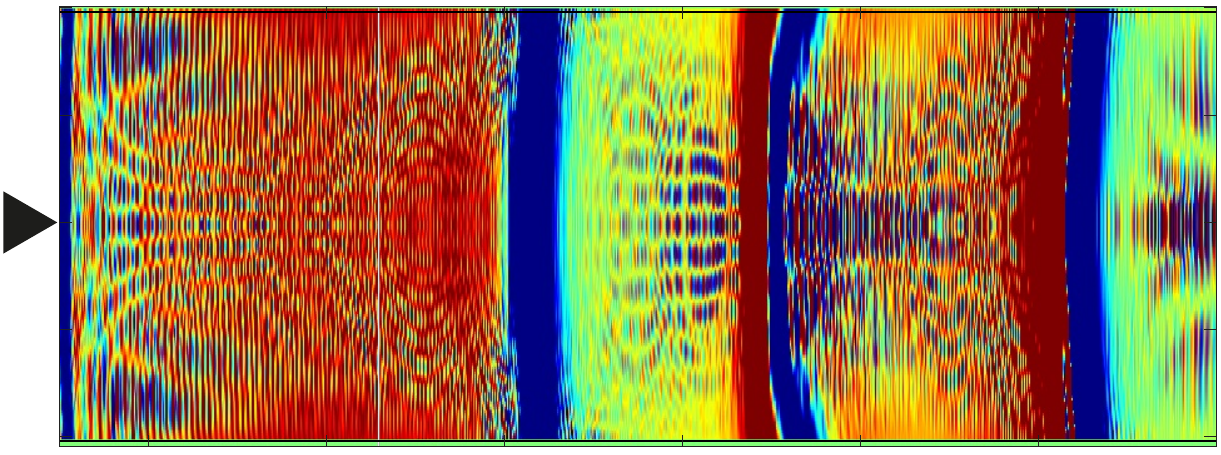}}
		\subfigure[]{\label{Fig:dcp8_rf_p4c}
			\includegraphics[width=1\columnwidth]{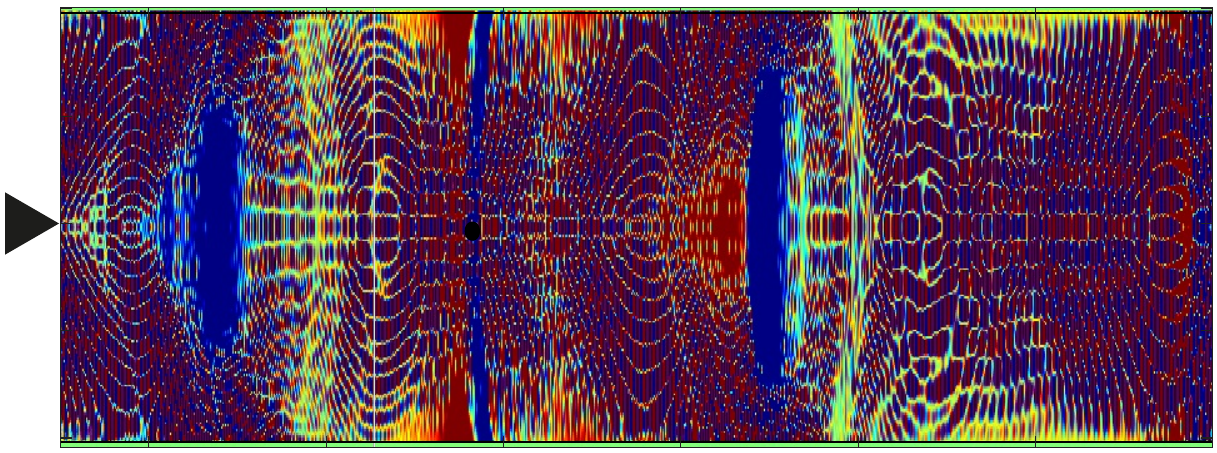}}
		\caption{Magnetic field distribution for magnetic flux excitation at frequency $\omega_\text{m}$ incident from the left, demonstrating frequency multiplication in a spatiotemporally modulated superconducting array with $\widetilde{\Phi}\text{dc}=0.8$. Panel (a) shows the field at $t=0.55$~ns with $\widetilde{\Phi}\text{rf}=0.2$. Panels (b)–(d)~correspond to $\widetilde{\Phi}\text{rf}=0.4$ at successive times: (b)~$t=0.5$~ns, (c)~$t=0.7$~ns, and (d)~$t=0.8$~ns.}
		\label{Fig:dcp8_rf_p4}
	\end{center}
\end{figure}

Figures~\ref{Fig:dcp8_rf_p2} to~\ref{Fig:dcp8_rf_p4c} further explore parameter-dependent harmonic development at fixed DC flux, systematically examining the role of the RF modulation amplitude $\widetilde{\Phi}_{\rm rf}$ while maintaining a constant static flux bias $\widetilde{\Phi}_{\rm dc}=0.8$. Figure~\ref{Fig:dcp8_rf_p2} shows the field distribution at $t=0.55$\,ns for a weak modulation $\widetilde{\Phi}_{\rm rf}=0.2$, revealing minimal wave distortion and predominantly fundamental-frequency propagation. Figures~\ref{Fig:dcp8_rf_p4a} to~\ref{Fig:dcp8_rf_p4c}, with increased modulation $\widetilde{\Phi}_{\rm rf}=0.4$, display the field at successive times $t=0.5, 0.7, 0.8$\,ns, demonstrating that stronger RF flux drive enhances the nonlinear wave mixing and leads to pronounced higher-harmonic content by $t=0.8$\,ns. These results provide a direct, controlled comparison that isolates the effect of modulation depth from DC bias. Together with Fig.~\ref{Fig:dcp6_rf_p6} ($\widetilde{\Phi}_{\rm dc}=\widetilde{\Phi}_{\rm rf}=0.6$, balanced harmonic generation) and Fig.~\ref{Fig:dcp4_rf_p85} ($\widetilde{\Phi}_{\rm dc}=0.4,\widetilde{\Phi}_{\rm rf}=0.85$, extreme broadband generation), Fig.~\ref{Fig:dcp8_rf_p4} indicates that the RF modulation amplitude is a critical, independent knob for controlling multiplication efficiency, enabling dynamic tuning of the harmonic output power without altering the DC operating point of the array. The time-domain simulations in Figs.~\ref{Fig:dcp6_rf_p6} to~\ref{Fig:dcp8_rf_p4} show the transient evolution of magnetic field distributions over sub-nanosecond timescales, chosen to resolve the fast nonlinear wave mixing and harmonic-generation dynamics inside the space-time-modulated bus. In contrast, actual qubit control pulses are typically $10$–$50$\,ns long, consistent with standard superconducting qubit gate durations.

\begin{figure}
	\begin{center}
			\includegraphics[width=1\columnwidth]{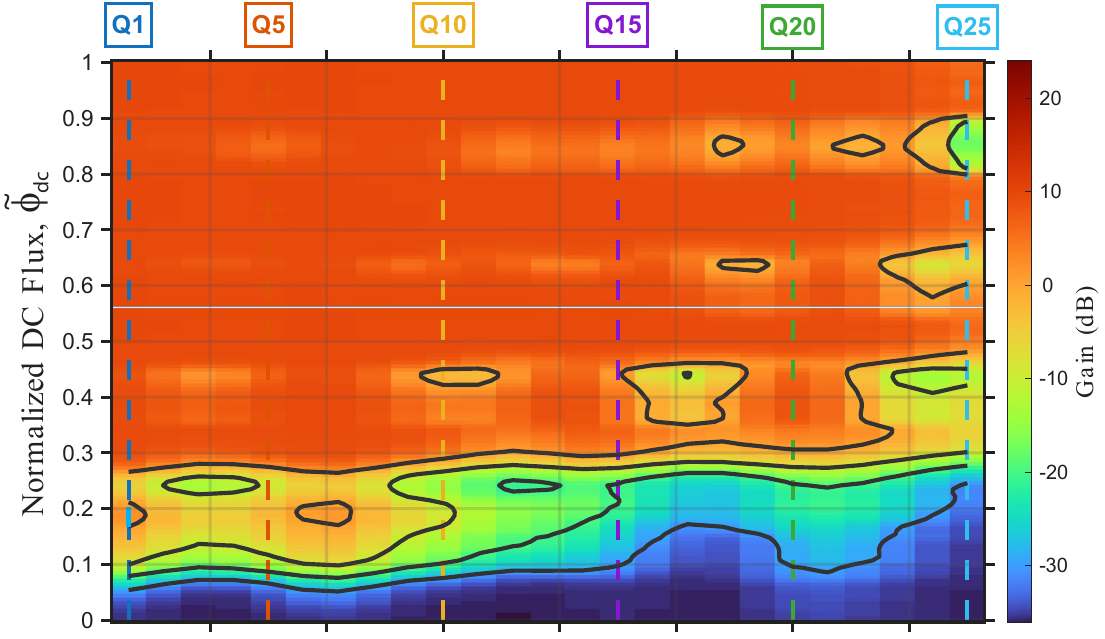}
		\caption{Selective excitation of different qubits in the circuit is achieved by tuning the DC flux bias $\widetilde{\Phi}_{\text{dc}}$, with the amplitude of the radio-frequency flux pulse fixed at $\widetilde{\Phi}_{\text{rf}}=0.85$ (simulation).}
		\label{Fig:dc}
	\end{center}
\end{figure}

\begin{figure}
	\begin{center}
		\includegraphics[width=1\columnwidth]{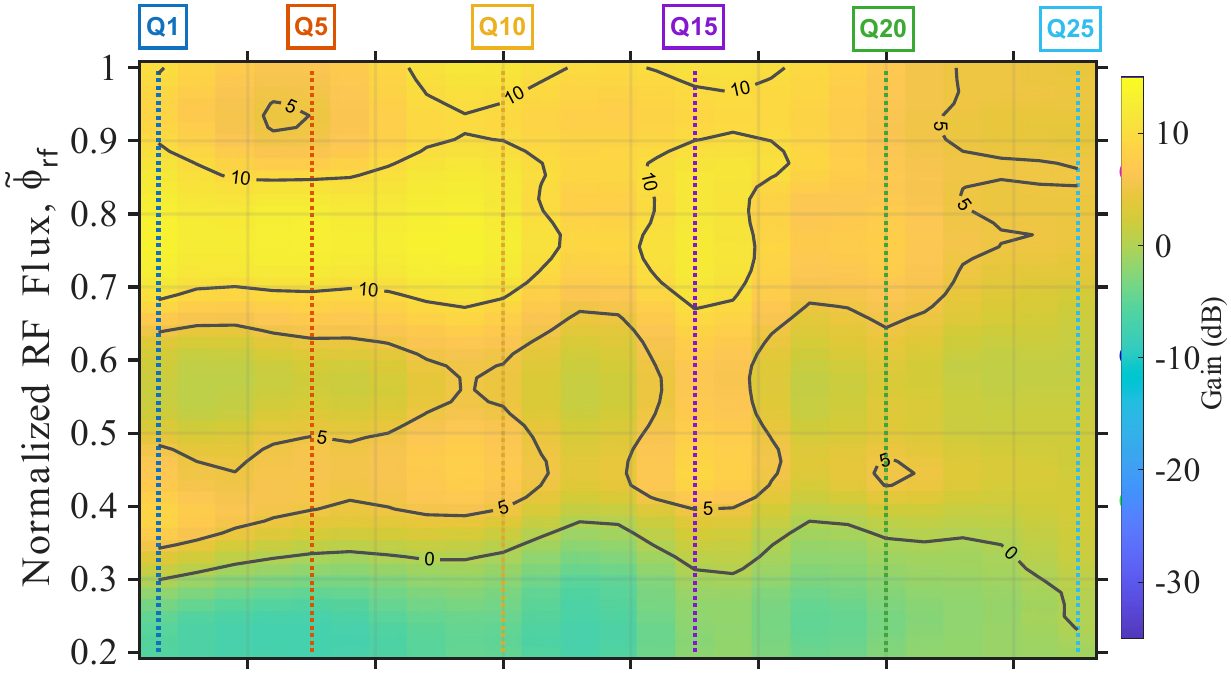}
		\caption{Probing qubit frequency tunability with fixed DC flux. The excitation spectra are simulated as a function of the normalized RF flux amplitude, \(\tilde{\Phi}_{\text{rf}}\), for a fixed DC flux bias of \(\tilde{\Phi}_{\text{dc}} = 0.8\). Distinct horizontal branches correspond to the flux-dependent transition frequencies of individual qubits (labeled Q1, Q5, ..., Q25).}
		\label{Fig:rf}
	\end{center}
\end{figure}

The individual addressability and frequency tunability of the qubits are shown in simulation in Fig.~\ref{Fig:dc}. By applying a fixed RF flux pulse amplitude ($\widetilde{\Phi}_{\rm rf}=0.85$) and systematically varying the DC flux bias $\widetilde{\Phi}_{\rm dc}$, distinct qubits in the array are selectively brought into resonance with the drive, each producing a well-separated resonance peak in the excitation spectrum. This indicates, in simulation, that the transition frequency of each qubit can be independently tuned over a significant range via its local flux bias — a prerequisite for targeted single-qubit addressing and for minimizing crosstalk in a multi-qubit architecture — and that a given qubit can be selected using the same fixed $\widetilde{\Phi}_{\rm rf}$ amplitude by adjusting only its DC working point, simplifying control-pulse sequencing. The clear separation between resonances suggests that, within the simulated model, qubits can be addressed individually without exciting their neighbors under these biasing conditions. These spectra are simulated. We note that Eq.~\eqref{eqa:L} requires $\widetilde{\Phi}_\text{dc}+\widetilde{\Phi}_\text{rf}<\pi/2$; for $\widetilde{\Phi}_\text{rf}=0.85$ this limits $\widetilde{\Phi}_\text{dc}$ to about $0.72$, and for $\widetilde{\Phi}_\text{dc}=0.8$ it limits $\widetilde{\Phi}_\text{rf}$ to about $0.77$. Features beyond these values (Figs.~\ref{Fig:dc} and~\ref{Fig:rf}) lie outside the strict validity range of the model and should be read as indicative only.

Complementary to Fig.~\ref{Fig:dc}, Fig.~\ref{Fig:rf} illustrates the frequency landscape of the array at a fixed DC flux bias of $\widetilde{\Phi}_{\rm dc}=0.8$ as the RF flux amplitude $\widetilde{\Phi}_{\rm rf}$ is varied. The resulting two-dimensional spectroscopy plot reveals distinct, quasi-horizontal branches, each tracing the resonant frequency of a specific qubit (e.g., Q1, Q5, Q10) as a function of the applied RF flux. While Fig.~\ref{Fig:dc} demonstrates selective excitation via DC bias at fixed drive amplitude, Fig.~\ref{Fig:rf} instead probes flux tunability and anharmonicity directly: the curvature of each branch reflects the qubit's deviation from a harmonic oscillator, with a flat branch indicating linear response and the observed curvature confirming the weak anharmonicity expected of a transmon. This characterization of the qubit Hamiltonian — quantifying flux tunability and anharmonicity — is useful for avoiding unwanted multi-level transitions during fast gate operations.

The modulation amplitudes $\tilde{\Phi}_{\rm dc}$ and $\tilde{\Phi}_{\rm rf}$ were selected based on the time-domain simulations of Figs.~\ref{Fig:dcp6_rf_p6} to~\ref{Fig:dcp8_rf_p4}, which show that these parameters provide independent control over harmonic-generation efficiency and spectral bandwidth. Moderate modulation depths (e.g., $\tilde{\Phi}_{\rm dc} = \tilde{\Phi}_{\rm rf} = 0.6$) yield balanced, well-separated harmonic generation suitable for addressing a discrete set of output tones, while stronger RF drive ($\tilde{\Phi}_{\rm rf} \gtrsim 0.85$) pushes the array into a broadband regime at the cost of increased back-reflection and nonlinear distortion. From a practical standpoint, only a single low-frequency pump tone is required to drive the entire multiplier, which reduces the number of high-frequency input lines and associated cryogenic filtering compared to architectures relying on multiple independently generated high-frequency drives. Achieving this in practice requires careful thermal engineering, since strong RF modulation of the Josephson array can increase quasiparticle generation; operating at moderate modulation amplitudes offers a reasonable balance between harmonic-generation efficiency and thermal load.

\begin{figure*}
	\begin{center}
		\includegraphics[width=1.8\columnwidth]{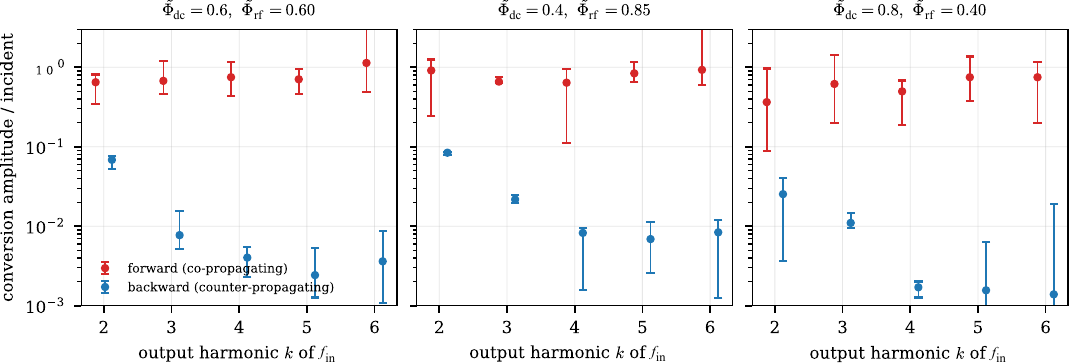}
		\caption{Forward and backward conversion amplitude, relative to the incident amplitude, in a one-dimensional space-time-modulated line model for three bias sets ($\widetilde{\Phi}_\text{dc},\widetilde{\Phi}_\text{rf}$ = 0.6, 0.6; 0.4, 0.85; 0.8, 0.4). Markers are medians and bars the range over four discretisations (cells per wavelength 24 and 48, Courant numbers 0.5 and 0.25).}
		\label{Fig:qconv}
	\end{center}
\end{figure*}

\subsection{Master-Equation Simulation of Multiplexed Qubit Control}\label{sec:qsim}
\begin{figure}
	\begin{center}
		\subfigure[]{\label{Fig:qgatea}
			\includegraphics[width=0.48\columnwidth]{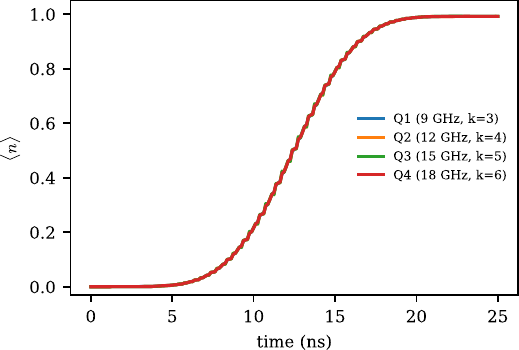}}
		\subfigure[]{\label{Fig:qgateb}
			\includegraphics[width=0.48\columnwidth]{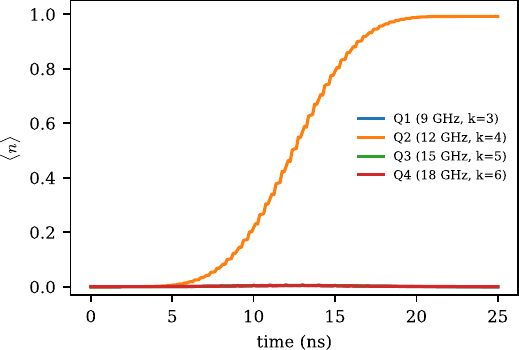}}
		\subfigure[]{\label{Fig:qgatec}
			\includegraphics[width=0.48\columnwidth]{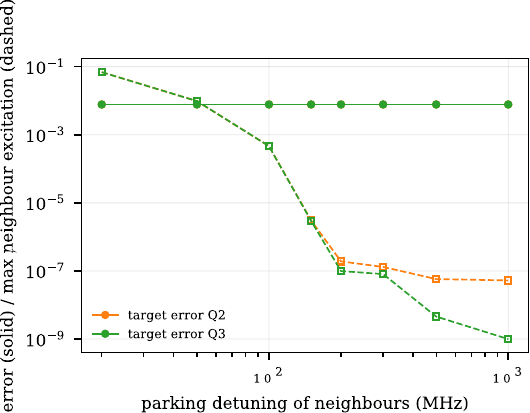}}
		\subfigure[]{\label{Fig:qgated}
			\includegraphics[width=0.48\columnwidth]{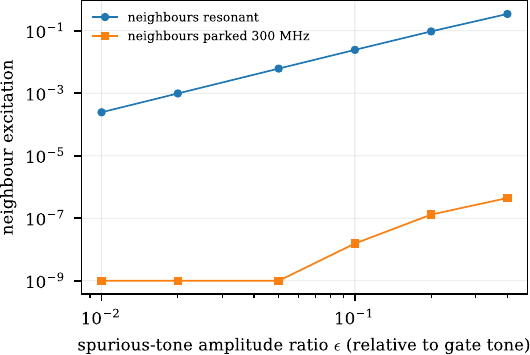}}
		\caption{Master-equation simulation of four transmons (9, 12, 15, 18 GHz) driven by harmonics 3 to 6 of a 3 GHz pump. (a) Population dynamics of a 25 ns $\pi$ pulse with all qubits resonant and equal tones. (b) Same tones with Q1, Q3 and Q4 parked 300 MHz away, so only Q2 is gated. (c) Target error (solid) and maximum neighbour excitation (dashed) versus parking detuning, with spurious tones at 20\% of the gate amplitude. (d) Neighbour excitation versus spurious-tone ratio $\epsilon$ for resonant and parked neighbours.}
		\label{Fig:qgate}
	\end{center}
\end{figure}

The full-wave results above are classical field simulations. To test the qubit-level claims, we simulate four flux-tunable transmons at $9$, $12$, $15$ and $18$\,GHz, addressed by harmonics $k=3$ to $6$ of a single $3$\,GHz pump (see the cryogenic measurement setup above), using a Lindblad master equation (QuTiP) with three levels per transmon, anharmonicity $\alpha/2\pi=-200$\,MHz, $T_1=T_\phi=50\,\mu$s and no rotating-wave approximation. The comb is treated as a semiclassical drive, and each tone amplitude is calibrated to produce a $\pi$ pulse in $T_\text{gate}=25$\,ns with a $\sin^2$ envelope, so the results depend on the relative, not the absolute, comb weights. Because the whole comb is present simultaneously, qubits at resonance rotate together (Fig.~\ref{Fig:qgate}a); individual gating is obtained by parking the non-target qubits, here $300$\,MHz away through their Z-lines (Fig.~\ref{Fig:qgate}b). The untuned $\pi$-pulse error of the target is about $0.8\%$ at $25$\,ns and falls to about $0.15\%$ at $100$\,ns, without pulse shaping such as DRAG. With spurious comb content at $20\%$ of the gate-tone amplitude, the excitation of parked neighbours falls from about $7\%$ at $20$\,MHz parking to below $10^{-3}$ at $100$\,MHz and beyond (Fig.~\ref{Fig:qgate}c). Spurious tones at $5\%$ of the gate amplitude excite resonant neighbours by about $0.6\%$, whereas parked neighbours are essentially unaffected (Fig.~\ref{Fig:qgate}d). These results support the use of a multiplexed comb together with Z-line parking for single-qubit gates, and they show that residual spectral purity of the comb, not the detuning of the parked qubits, sets the crosstalk floor. Leakage out of the computational subspace is a known risk of multi-tone drive, where spurious tones near the 1-2 or multiphoton transitions cause errors~\cite{shi2023multiplexed}. In our simulation the level-2 population of the gated qubit after the $25$\,ns pulse is about $10^{-5}$, that of parked neighbours is below $10^{-7}$, and adding further comb tones at $k=1,2,7,8$ with amplitudes equal to the gate tone leaves both unchanged. The $0.8\%$ error is therefore not leakage. The $3$\,GHz tone spacing is much larger than $|\alpha|$, which keeps single-photon leakage far off resonance, but the sum of two comb tones can fall within about $|\alpha|$ of a two-photon transition, so a calibrated comb should be checked for such coincidences. As an alternative to parking, pulse-shape-invariant compilation, in which all qubits receive identical pulse envelopes and arbitrary gates are built from pairs of $\pi/2$ pulses and virtual $Z$ gates~\cite{shi2023multiplexed}, appears naturally compatible with a comb that inherits a common envelope from the pump, provided the per-qubit drive strengths are equalized.

The directional asymmetry $\kappa_i$ is estimated with a one-dimensional space-time-modulated line model (flux-based telegrapher equations with the inductance of Eq.~\eqref{eqa:L}, a co-propagating modulation, and a discrete ladder). Fig.~\ref{Fig:qconv} shows forward and backward conversion for three bias sets, with the spread over four discretisations. Here $f_\text{in}=\omega_\text{m}/2\pi$ is the frequency of the incident tone and $k$ is the harmonic order, so the output frequency is $k f_\text{in}$. The backward conversion is small and robust, about $0.05$--$0.09$ of the incident amplitude at $k=2$ and below $0.03$ for $k\geq3$, whereas the forward conversion is large but depends on the assumed ladder cell size. The ratio of backward to forward conversion is therefore only an order-of-magnitude estimate, $\kappa\lesssim0.3$ at $k=2$ and $\kappa\lesssim0.06$ for $k\geq3$, which includes the harmonics addressing the qubits here. The nondispersive assumption is thus optimistic, and a quantitative $\kappa_i$ requires the full-wave model of the fabricated bus or measurement.

These simulations are semiclassical, use assumed qubit parameters and a frequency-independent drive coupling, and are not experimental data. The qubit frequencies considered here ($9$ to $18$\,GHz) lie in the microwave band; operation at millimeter-wave frequencies is not demonstrated in this work.

\section{Conclusions}\label{sec:conc} 
We presented an on-chip nonreciprocal space-time-periodic Josephson frequency multiplier that up-converts a single low-frequency input into a comb of high-order harmonics while enforcing directional signal propagation along the bus. Full-wave time-domain simulations showed that the harmonic content and bandwidth of the generated comb can be tailored through the applied DC and RF flux bias, and that individual output channels can be selectively addressed by tuning the bus operating point. The nonreciprocal nature of the bus further provides directional isolation between channels, offering a route to suppressing back-reflection and inter-channel coupling without additional off-chip components. By combining frequency multiplexing, directional isolation, and parametric frequency conversion within a single compact element, the proposed device offers a promising building block for cryogenic microwave systems (and, in principle, millimeter-wave systems) requiring multiplexed delivery of tones to spatially distributed, frequency-addressable loads, including the potential control of multi-qubit superconducting arrays. Master-equation simulations of a four-qubit array indicate that, with Z-line parking, single-qubit gates can be addressed individually with low crosstalk, while a one-dimensional line model indicates a small backward conversion; both are simulation results and experimental validation remains to be done. These results establish space-time-periodic modulation and engineered nonreciprocity as promising design principles for compact, on-chip frequency-multiplexed control in next-generation cryogenic microwave systems.

\vspace{0.5cm}
\textbf{Conﬂict of Interest}

The authors declare no conﬂict of interest.

\vspace{0.5cm}
\textbf{Data Availability Statement}

The data that support the ﬁndings of this study are available from the cor-
responding author upon reasonable request.

\vspace{0.5cm}
\textbf{Keywords}

Superconducting qubits, Josephson junctions, harmonic-generation, space-time modulation, nonreciprocity.

\bibliographystyle{IEEEtran}
\bibliography{Taravati_Reference.bib}

% Generated by IEEEtran.bst, version: 1.14 (2015/08/26)
\begin{thebibliography}{10}
\providecommand{\url}[1]{#1}
\csname url@samestyle\endcsname
\providecommand{\newblock}{\relax}
\providecommand{\bibinfo}[2]{#2}
\providecommand{\BIBentrySTDinterwordspacing}{\spaceskip=0pt\relax}
\providecommand{\BIBentryALTinterwordstretchfactor}{4}
\providecommand{\BIBentryALTinterwordspacing}{\spaceskip=\fontdimen2\font plus
\BIBentryALTinterwordstretchfactor\fontdimen3\font minus
  \fontdimen4\font\relax}
\providecommand{\BIBforeignlanguage}[2]{{%
\expandafter\ifx\csname l@#1\endcsname\relax
\typeout{** WARNING: IEEEtran.bst: No hyphenation pattern has been}%
\typeout{** loaded for the language `#1'. Using the pattern for}%
\typeout{** the default language instead.}%
\else
\language=\csname l@#1\endcsname
\fi
#2}}
\providecommand{\BIBdecl}{\relax}
\BIBdecl

\bibitem{berlin2020axion}
A.~Berlin, R.~T. D’Agnolo, S.~A. Ellis, C.~Nantista, J.~Neilson, P.~Schuster,
  S.~Tantawi, N.~Toro, and K.~Zhou, ``Axion dark matter detection by
  superconducting resonant frequency conversion,'' \emph{Journal of High Energy
  Physics}, vol. 2020, no.~7, pp. 1--42, 2020.

\bibitem{dixit2021searching}
A.~V. Dixit, S.~Chakram, K.~He, A.~Agrawal, R.~K. Naik, D.~I. Schuster, and
  A.~Chou, ``Searching for dark matter with a superconducting qubit,''
  \emph{{Phys. Rev. Lett.}}, vol. 126, no.~14, p. 141302, 2021.

\bibitem{bass2024quantum}
S.~D. Bass and M.~Doser, ``Quantum sensing for particle physics,'' \emph{Nat.
  Rev. Phys.}, pp. 1--11, 2024.

\bibitem{han2021microwave}
X.~Han, W.~Fu, C.-L. Zou, L.~Jiang, and H.~X. Tang, ``Microwave-optical quantum
  frequency conversion,'' \emph{Optica}, vol.~8, no.~8, pp. 1050--1064, 2021.

\bibitem{albrecht2014waveguide}
B.~Albrecht, P.~Farrera, X.~Fernandez-Gonzalvo, M.~Cristiani, and
  H.~De~Riedmatten, ``A waveguide frequency converter connecting rubidium-based
  quantum memories to the telecom c-band,'' \emph{{Nat. Commun.}}, vol.~5,
  no.~1, p. 3376, 2014.

\bibitem{bock2018high}
M.~Bock, P.~Eich, S.~Kucera, M.~Kreis, A.~Lenhard, C.~Becher, and J.~Eschner,
  ``High-fidelity entanglement between a trapped ion and a telecom photon via
  quantum frequency conversion,'' \emph{{Nat. Commun.}}, vol.~9, no.~1, p.
  1998, 2018.

\bibitem{maring2018quantum}
N.~Maring, D.~Lago-Rivera, A.~Lenhard, G.~Heinze, and H.~de~Riedmatten,
  ``Quantum frequency conversion of memory-compatible single photons from 606
  nm to the telecom c-band,'' \emph{Optica}, vol.~5, no.~5, pp. 507--513, 2018.

\bibitem{santiago2022resonant}
T.~Santiago-Cruz, S.~D. Gennaro, O.~Mitrofanov, S.~Addamane, J.~Reno,
  I.~Brener, and M.~V. Chekhova, ``Resonant metasurfaces for generating complex
  quantum states,'' \emph{Science}, vol. 377, no. 6609, pp. 991--995, 2022.

\bibitem{taravati2024efficient}
S.~Taravati, ``Efficient nonreciprocal frequency conversion with space-time
  {J}osephson junction metasurfaces,'' in \emph{2024 54th European Microwave
  Conference (EuMC)}.\hskip 1em plus 0.5em minus 0.4em\relax IEEE, 2024, pp.
  600--603.

\bibitem{geus2024low}
J.~F. Geus, F.~Elsen, S.~Nyga, A.~J. Stolk, K.~L. van~der Enden, E.~J. van
  Zwet, C.~Haefner, R.~Hanson, and B.~Jungbluth, ``Low-noise short-wavelength
  pumped frequency downconversion for quantum frequency converters,''
  \emph{Optica Quantum}, vol.~2, no.~3, pp. 189--195, 2024.

\bibitem{gambetta2017building}
J.~M. Gambetta, J.~M. Chow, and M.~Steffen, ``Building logical qubits in a
  superconducting quantum computing system,'' \emph{npj quantum information},
  vol.~3, no.~1, p.~2, 2017.

\bibitem{he2022control}
Y.~He, J.~Liu, C.~Zhao, R.~Huang, G.~Dai, and W.~Chen, ``Control system of
  superconducting quantum computers,'' \emph{Journal of Superconductivity and
  Novel Magnetism}, vol.~35, no.~1, pp. 11--31, 2022.

\bibitem{clarke2008superconducting}
J.~Clarke and F.~K. Wilhelm, ``Superconducting quantum bits,'' \emph{Nature},
  vol. 453, no. 7198, pp. 1031--1042, 2008.

\bibitem{kjaergaard2020superconducting}
M.~Kjaergaard, M.~E. Schwartz, J.~Braum{\"u}ller, P.~Krantz, J.~I.-J. Wang,
  S.~Gustavsson, and W.~D. Oliver, ``Superconducting qubits: Current state of
  play,'' \emph{Annual Review of Condensed Matter Physics}, vol.~11, no.~1, pp.
  369--395, 2020.

\bibitem{anferov2024superconducting}
A.~Anferov, S.~P. Harvey, F.~Wan, J.~Simon, and D.~I. Schuster,
  ``Superconducting qubits above 20 {GH}z operating over 200 m{K},'' \emph{PRX
  Quantum}, vol.~5, no.~3, p. 030347, 2024.

\bibitem{acharya2023multiplexed}
R.~Acharya, S.~Brebels, A.~Grill, J.~Verjauw, T.~Ivanov, D.~P. Lozano, D.~Wan,
  J.~Van~Damme, A.~Vadiraj, M.~Mongillo \emph{et~al.}, ``Multiplexed
  superconducting qubit control at millikelvin temperatures with a low-power
  cryo-cmos multiplexer,'' \emph{Nature Electronics}, vol.~6, no.~11, pp.
  900--909, 2023.

\bibitem{forgione2004enhancements}
J.~B. Forgione, D.~J. Benford, E.~D. Buchanan, S.~H. Moseley~Jr, J.~Rebar, and
  R.~A. Shafer, ``Enhancements to a superconducting quantum interference device
  (squid) multiplexer readout and control system,'' in \emph{Millimeter and
  Submillimeter Detectors for Astronomy II}, vol. 5498.\hskip 1em plus 0.5em
  minus 0.4em\relax SPIE, 2004, pp. 784--795.

\bibitem{shi2023multiplexed}
P.~Shi, J.~Yuan, F.~Yan, and H.~Yu, ``Multiplexed control scheme for scalable
  quantum information processing with superconducting qubits,'' \emph{arXiv
  preprint arXiv:2312.06911}, 2023.

\bibitem{takeuchi2024microwave}
N.~Takeuchi, T.~Yamae, T.~Yamashita, T.~Yamamoto, and N.~Yoshikawa,
  ``Microwave-multiplexed qubit controller using adiabatic superconductor
  logic,'' \emph{npj Quantum Information}, vol.~10, no.~1, p.~53, 2024.

\bibitem{zhao2024multiplexed}
P.~Zhao, ``A multiplexed control architecture for superconducting qubits with
  row-column addressing,'' \emph{arXiv preprint arXiv:2403.03717}, 2024.

\bibitem{anferov2025millimeter}
A.~Anferov, F.~Wan, S.~P. Harvey, J.~Simon, and D.~I. Schuster,
  ``Millimeter-wave superconducting qubit,'' \emph{PRX Quantum}, vol.~6, no.~2,
  p. 020336, 2025.

\bibitem{Taravati_Kishk_MicMag_2019}
S.~Taravati and A.~A. Kishk, ``Space-time modulation: Principles and
  applications,'' \emph{IEEE Microw. Mag.}, vol.~21, no.~4, pp. 30--56, 2020.

\bibitem{Taravati_ACSP_2022}
S.~Taravati and G.~V. Eleftheriades, ``Microwave space-time-modulated
  metasurfaces,'' \emph{{ACS Photonics}}, vol.~9, no.~2, pp. 305--318, 2022.

\bibitem{taravati2024spatiotemporal}
S.~Taravati, ``Photon-blockade analogue nonreciprocal absorption in
  spatiotemporal metasurfaces,'' \emph{arXiv preprint arXiv:2409.08137}, 2024.

\bibitem{taravati20234d}
S.~Taravati and G.~V. Eleftheriades, ``4{D} wave transformations enabled by
  space-time metasurfaces: Foundations and illustrative examples,'' \emph{IEEE
  Antennas Propag. Mag.}, vol.~65, no.~4, pp. 61--74, 2023.

\bibitem{Taravati_Kishk_PRB_2018}
S.~Taravati and A.~A. Kishk, ``Dynamic modulation yields one-way beam
  splitting,'' \emph{{Phys. Rev. B}}, vol.~99, no.~7, p. 075101, Jan. 2019.

\bibitem{zhang2021magnetic}
D.~Zhang and J.-S. Tsai, ``Magnetic-free traveling-wave nonreciprocal
  superconducting microwave components,'' \emph{{Phys. Rev. Appl.}}, vol.~15,
  no.~6, p. 064013, 2021.

\bibitem{taravati2025light}
S.~Taravati, ``Light transmission through space-time-modulated {J}osephson
  junction arrays and application to quantum angular-frequency beam
  multiplexing,'' \emph{{IEEE Trans. Antennas Propagat.}}, 2025.

\bibitem{zhuang2024superconducting}
Y.~Zhuang, C.~Gaikwad, D.~Kowsari, K.~Murch, and A.~Nagulu, ``Superconducting
  isolators based on time-modulated coupled-resonator systems,'' \emph{{Phys.
  Rev. Appl.}}, vol.~21, no.~5, p. 054061, 2024.

\bibitem{metelmann2015nonreciprocal}
A.~Metelmann and A.~A. Clerk, ``Nonreciprocal photon transmission and
  amplification via reservoir engineering,'' \emph{Phys. Rev. X}, vol.~5,
  no.~2, p. 021025, 2015.

\bibitem{Taravati_PRAp_2018}
S.~Taravati, ``Giant linear nonreciprocity, zero reflection, and zero band gap
  in equilibrated space-time-varying media,'' \emph{Phys. Rev. Appl.}, vol.~9,
  no.~6, p. 064012, Jun. 2018.

\bibitem{taravati2020full}
S.~Taravati and G.~V. Eleftheriades, ``Full-duplex nonreciprocal beam steering
  by time-modulated phase-gradient metasurfaces,'' \emph{{Phys. Rev. Appl.}},
  vol.~14, no.~1, p. 014027, 2020.

\bibitem{taravati2025_entangle}
S.~Taravati, ``Nonreciprocal entanglement of frequency-distinct qubits,''
  \emph{Advanced Quantum Technologies}, vol.~8, no.~10, p. e2500171, 2025.

\bibitem{Taravati_AMA_PRApp_2020}
S.~Taravati and G.~V. Eleftheriades, ``Space-time medium functions as a perfect
  antenna-mixer-amplifier transceiver,'' \emph{{Phys. Rev. Appl.}}, vol.~14,
  no.~5, p. 054017, 2020.

\bibitem{taravati2025finite}
S.~Taravati, A.~A. Kishk, and G.~V. Eleftheriades, ``Finite-difference
  time-domain simulation of wave propagation in space-time-varying media:
  Derivation of schemes, extended framework, and illustrative examples,''
  \emph{IEEE Antennas and Propagation Magazine}, 2025.

\bibitem{taravati2025designing}
S.~Taravati, ``Designing space-time metamaterials: The central role of
  dispersion engineering,'' \emph{arXiv preprint arXiv:2511.19541}, 2025.

\bibitem{taravati2021pure}
S.~Taravati and G.~V. Eleftheriades, ``Pure and linear frequency-conversion
  temporal metasurface,'' \emph{{Phys. Rev. Appl.}}, vol.~15, no.~6, p. 064011,
  2021.

\bibitem{Taravati_PRB_Mixer_2018}
S.~Taravati, ``Aperiodic space-time modulation for pure frequency mixing,''
  \emph{{Phys. Rev. B}}, vol.~97, no.~11, p. 115131, 2018.

\bibitem{taravati_PRApp_2019}
S.~Taravati and G.~V. Eleftheriades, ``Generalized space-time periodic
  diffraction gratings: Theory and applications,'' \emph{{Phys. Rev. Appl.}},
  vol.~12, no.~2, p. 024026, 2019.

\bibitem{taravati2024one}
S.~Taravati, ``One-way absorption and isolation in space-time-periodic
  superconducting metasurfaces,'' in \emph{2024 Eighteenth International
  Congress on Artificial Materials for Novel Wave Phenomena
  (Metamaterials)}.\hskip 1em plus 0.5em minus 0.4em\relax IEEE, 2024, pp.
  1--3.

\bibitem{GamowUniverse}
G.~Gamow, ``The evolution of the universe,'' \emph{Nature}, vol. 162, no. 4122,
  1948.

\bibitem{ginis2010frequency}
V.~Ginis, P.~Tassin, B.~Craps, and I.~Veretennicoff, ``Frequency converter
  implementing an optical analogue of the cosmological redshift,'' \emph{{Opt.
  Expr.}}, vol.~18, no.~5, pp. 5350--5355, 2010.

\bibitem{westerberg2014experimental}
N.~Westerberg, S.~Cacciatori, F.~Belgiorno, F.~D. Piazza, and D.~Faccio,
  ``Experimental quantum cosmology in time-dependent optical media,'' \emph{New
  Journal of Physics}, vol.~16, no.~7, p. 075003, 2014.

\end{thebibliography}

\end{document}